\documentclass[iop]{emulateapj}
\usepackage{amsmath}
\usepackage{color}

\newcommand{\Chandra}{\textit{Chandra}}
\newcommand{\XMM}{XMM-\textit{Newton}}
\newcommand{\ROSAT}{\textit{ROSAT}}

\begin{document}

\slugcomment{accepted by ApJ}
\shorttitle{HOT ACCRETED HALO OF NGC 891}
\shortauthors{HODGES-KLUCK, BREGMAN, AND LI}

\title{The Hot, Accreted Halo of NGC 891}

\author{Edmund J. Hodges-Kluck$^{1}$, Joel~N. Bregman$^{1}$ \&
  Jiang-tao Li$^{1}$}
\altaffiltext{1}{Department of Astronomy, University of Michigan, Ann
  Arbor, MI 48109}
\email{hodgeskl@umich.edu}

\begin{abstract}
Galaxies are surrounded by halos of hot gas whose mass and origin remain unknown. One of the most challenging properties to measure is the metallicity, which constrains both of these. We present a measurement of the metallicity around NGC~891, a nearby, edge-on, Milky Way analog. We find that the hot gas is dominated by low metallicity gas near the virial temperature at $kT=0.20\pm0.01$~keV and $Z/Z_{\odot} = 0.14\pm0.03$(stat)$^{+0.08}_{-0.02}$(sys), and that this gas co-exists with hotter ($kT=0.71\pm0.04$~keV) gas that is concentrated near the star-forming regions in the disk. Model choices lead to differences of $\Delta Z/Z_{\odot} \sim 0.05$, and higher $S/N$ observations would be limited by systematic error and plasma emission model or abundance ratio choices. 
The low metallicity gas is consistent with the inner part of an extended halo accreted from the intergalactic medium, which has been modulated by star formation. However, there is much more cold gas than hot gas around NGC~891, which is difficult to explain in either the accretion or supernova-driven outflow scenarios. 
We also find a diffuse nonthermal excess centered on the galactic center and extending to 5~kpc above the disk with a 0.3-10~keV $L_X = 3.1\times 10^{39}$~erg~s$^{-1}$. This emission is inconsistent with inverse Compton scattering or single-population synchrotron emission, and its origin remains unclear. 
\end{abstract}


\section{Introduction}
\label{section.intro}

A basic prediction of $\Lambda$CDM galaxy-formation models is that galaxies at or above the \citet{schechter76} $L_*$ are surrounded by massive reservoirs of hot gas at $T>10^6$~K \citep{white78,white91,crain10}. Galaxies form from gas that flows into a dark-matter potential, and much of the gravitational energy is converted to heat at an accretion shock, which heats the gas to near the virial temperature. Above $T \sim 10^6$~K the dynamical time exceeds the cooling time, so galaxies massive enough for the virial temperature to fall in this regime ($v_{\text{rot}} \gtrsim 100$~km~s$^{-1}$) are surrounded by massive, quasi-hydrostatic hot halos of primordial gas. These hot halos would provide most of the fuel for long-term star formation at $z<1$ and mediate inflows and outflows. 

There is now compelling evidence that these halos exist around super-$L_*$ galaxies \citep{anderson11,humphrey11,dai12,bogdan13,anderson16,li17,bogdan17}. Meanwhile, both emission- and absorption-studies of the Milky Way's hot halo indicate that it is extended and potentially very massive \citep{gupta14,miller15,nicastro16,faerman17,bregman18}. However, there is limited evidence for extended halos around other individual $L_*$ galaxies, although \citet{anderson13} showed that hot gas is ubiquitous around these galaxies within $r<50$~kpc by stacking \ROSAT{} data (albeit with a lower luminosity than in the massive galaxies). 

Instead, shallow X-ray images revealed that, while extraplanar hot gas is ubiquitous around edge-on galaxies, these X-ray coronae usually look like thick disks or follow galactic outflows seen at other wavelengths, and have luminosities correlated with the star-formation rate (SFR) of the galaxy \citep{strickland04,tullmann06,li13a}. This strongly implicates galactic feedback as the origin of this hot gas, and since hot outflows will virialize in the dark-matter halo in a similar manner to accreted infall, it is not yet clear whether $L_*$ galaxies generally have extended hot halos and how much of the mass was built up by infall. 

In general, we do not understand how the hot gas participates in the disk-halo cycle, for which we need the temperature, metallicity, and kinematic information. The metallicity is a particularly important measurement because it can distinguish enriched feedback ejecta from accreted intergalactic material, although an important caveat is that X-ray faint elliptical galaxies and the bulges of some spiral galaxies have low metallicity hot ISM \citep{humphrey06,tang09,li_zhiyuan11} that is unlikely to result from accreted infall \citep{su13}. Early- and late-type galaxies may have substantially different halo properties and origins. The temperature and metallicity can be measured at the energy resolution of the \Chandra{} ACIS or \XMM{} EPIC CCDs, but at low $S/N$ the simple thermal models that often fit these spectra can be misleading, and one cannot perform a spatially resolved analysis to separate gas that originated in the galaxy from more primordial material. Thus, existing measurements of the metallicity around massive spirals that indicate that the gas is mostly accreted are not necessarily reliable \citep[e.g.,][]{anderson16,bogdan17}.

\begin{figure*}
\begin{center}
\includegraphics[width=1\textwidth]{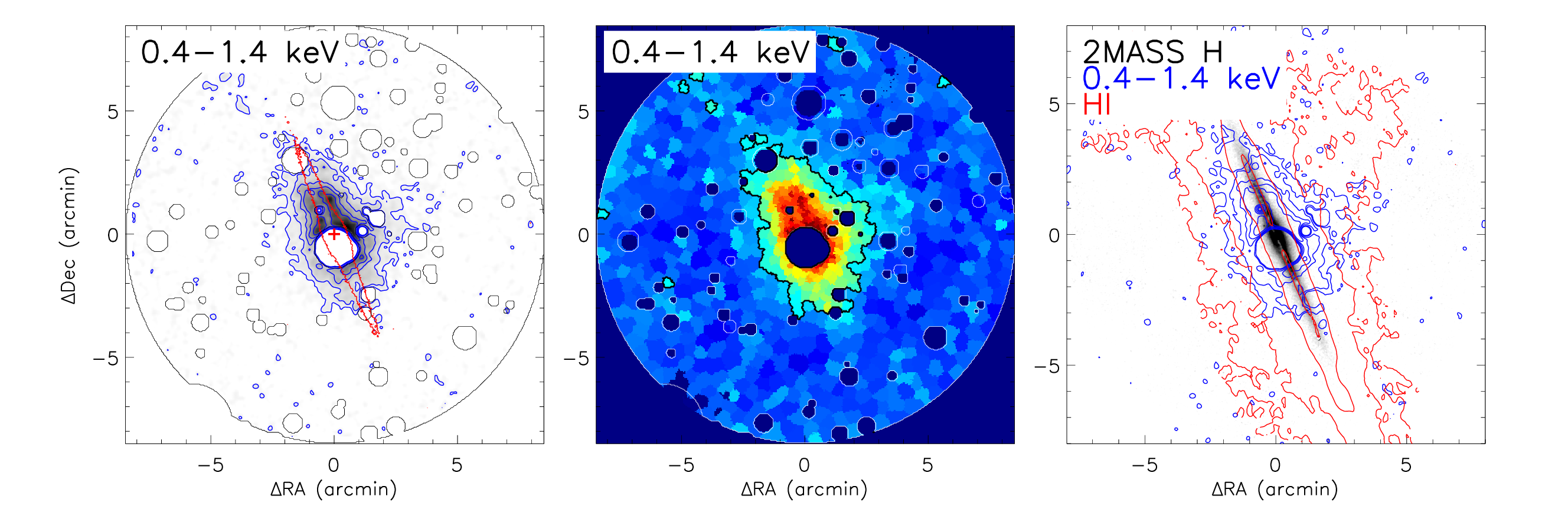}
\includegraphics[width=1\textwidth]{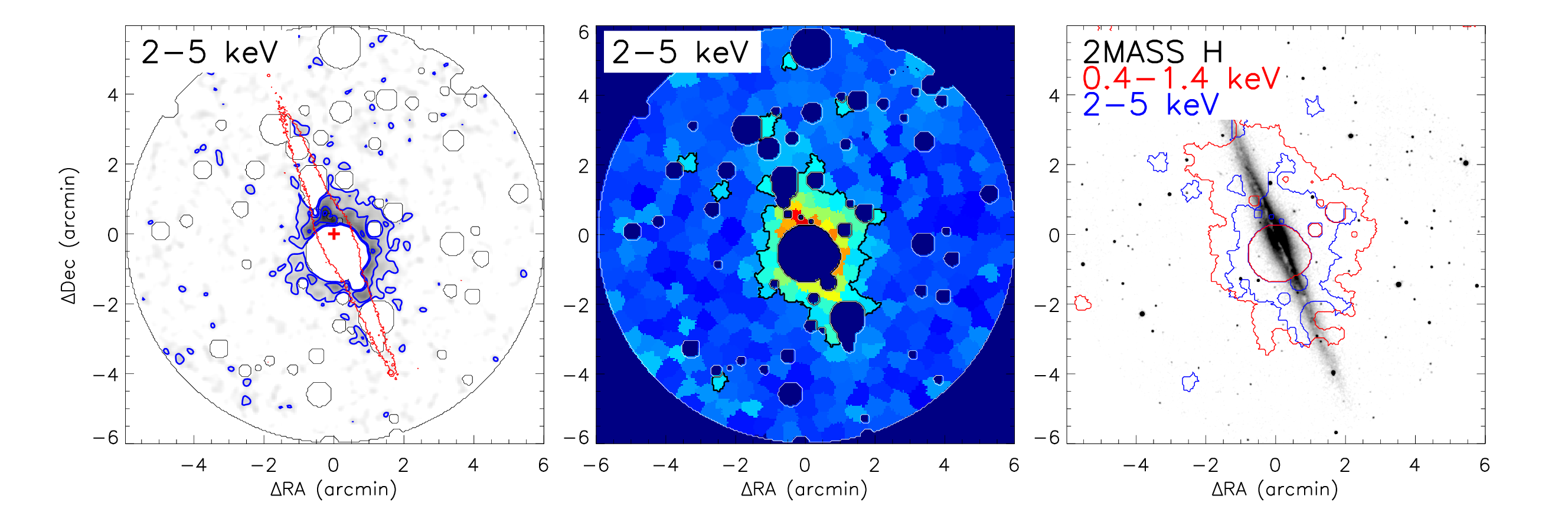}
\caption{Images from the combined 260\,ks (MOS and pn) data in the soft (top) and hard (bottom) bands. The first column shows the X-ray emission in grayscale, blue contours at the 2, 4, 8, and 12$\sigma$ level above background, and the 2MASS $H$-band disk of the galaxy. The small black circles are point source masks, and the image is clipped at the large black circle (outside of which the effective exposure is lower). The central column shows an adaptively binned image with the average intensity per bin in rainbow color and the 3$\sigma$ contour above background as a dark black line. The right-hand column shows the X-ray contours overlaid on the 2MASS $H$-band image. The top-right panel also shows H{\sc i} contours at 1, 10, 100, and 1000$\times 10^{19}$~cm$^{-2}$ in red, while the bottom-right panel shows the 3$\sigma$ contours above background from the adaptively binned images in the soft (red) and hard (blue) bands to show the spatial correspondence.}
\label{figure.xmm_images}
\end{center}
\end{figure*}

Here we address the temperature,  metallicity ($Z/Z_{\odot}$), and cooling rate in the inner region of a hot halo, using deep \XMM{} data for NGC~891, which is a nearby \citep[$d=10$\,Mpc;][]{temple05}, edge-on, Milky Way analog with the brightest X-ray halo among normal galaxies \citep[the \textit{luminosity} of a few$\times 10^{39}$\,erg\,s$^{-1}$ is normal for its size;][]{anderson13}. At this distance, 1\,arcmin corresponds to about 2.7\,kpc. NGC~891 has an inclination of 89.5$^{\circ}$ \citep{rupen91}, which is ideal for separating the disk and halo gas. It is also very similar to the Milky Way, with an apparent diameter ($D_{25}$) of 40\,kpc, a rotation velocity of about 210\,km\,s$^{-1}$ \citep{oosterloo07}, a spectral type of SBb \citep{garcia-burillo95}, and a $B$-band luminosity of $L_B = 2.6\times 10^{10} L_{\odot,B}$ \citep{devaucouleurs91}, although its SFR$= 3.8 M_{\odot}$\,yr$^{-1}$ is somewhat higher than in the Milky Way \citep{popescu04}. Finally, NGC~891 is surrounded by a massive \ion{H}{1} halo \citep[$1.2 \times 10^9 M_{\odot}$;][]{oosterloo07}, whose interaction with the hot halo remains unclear.  

The remainder of this paper is organized as follows: In Section~\ref{section.obs}, we describe the observations and reduction. In Section~\ref{section.images}, we use the soft and hard X-ray images from combining the data to characterize the diffuse halo morphology, and in Section~\ref{section.spectra} we extract spectra from the halo and fit them to obtain the temperature, metallicity, and mass of the hot gas. In this section we also examine the possible sources of bias in these measurements. In Section~\ref{section.discussion} we argue that most of the hot gas is from an accreted hot halo, that prior fits to lower signal spectra around other late-type galaxies are essentially accurate, that the galactic fountain model has problems balancing the amount of hot and cold gas in NGC~891, and that there is a diffuse, nonthermal X-ray emision component of unknown origin. We summarize our results and conclusions in Section~\ref{section.summary}.

\section{Observations}
\label{section.obs}

\begin{deluxetable*}{ccc|cccc|cccc}
\tablenum{1}
\label{table.obs}
\tabletypesize{\scriptsize}
\tablecaption{NGC 891 Observations}
\tablewidth{0pt}
\tablehead{
\colhead{} & \colhead{} & \colhead{} & \multicolumn{4}{c}{$t_{\text{exp}}$ (ks)} & \multicolumn{4}{c}{$\Sigma$GTI (ks)} \\
\colhead{Date} & \colhead{Telescope} & \colhead{ObsID} & \colhead{ACIS-S} & \colhead{pn} & \colhead{MOS1} & \colhead{MOS2} & \colhead{ACIS-S} & \colhead{pn} & \colhead{MOS1} & \colhead{MOS2}  
}
\startdata
2000-11-01 & CXO & 794 & 50.9 & 	& & & 29.7 & & & \\ 
2003-12-10 & CXO & 4613 & 118.8 & & & & 111.0 & & & \\
2011-08-25 & XMM & 0670950101 &  & 126.1 & 126.2 & 110.5 & & 97.8 & 103.8 & 105.0 \\
2017-01-27 & XMM & 0780760101 &  & 54.2 & 69.5 & 69.6 & & 22.4 & 27.2 & 27.9 \\
2017-01-29 & XMM & 0780760201 &  & 56.7 & 70.6 & 70.7 & & 32.5 & 36.1 & 38.8 \\
2017-02-19 & XMM & 0780760401 &  & 60.0 & 72.0 & 72.1 & & 30.1 & 34.6 & 35.0 \\
2017-02-23 & XMM & 0780760301 &  & 55.7 & 70.9 & 71.2 & & 21.7 & 26.9 & 28.2 \\
2017-02-25 & XMM & 0780760501 &  & 61.3 & 71.4 & 71.5 & & 34.1 & 48.2 & 54.5 \\
\hline
Total		& 				& & 169.7 & 414.0 & 480.6 & 465.6 & 140.7 & 216.2 & 276.8 & 289.4
\enddata
\end{deluxetable*}

\subsection{Data Preparation}

We used X-ray data from \XMM{} and \Chandra{}. The observations are summarized in Table~\ref{table.obs}, of which the \Chandra{} and 2011 \XMM{} observation were published in our previous study \citep{hk13}, while \citet{strickland04} and \citet{li13a} also examined the extraplanar hot gas in the \Chandra{} data. The 2017 \XMM{} data was obtained through our GO program \#078076. 

We processed the data using standard methods. For the \XMM{} data we used the Scientific Analysis System (SAS v16.0.0) software, including the extended source analysis software \citep[ESAS;][]{snowden04}. We used the SAS tools {\tt emchain} and {\tt epchain} to register the events for the MOS and pn cameras, then the ESAS {\tt mos-filter} and {\tt pn-filter} tools to exclude periods where there is significant contamination from soft proton flares. The filter scripts produce a count-rate histogram in bins of 100--200~s to identify flaring periods, so there can be multiple good-time intervals per exposure. The total resulting good-time intervals are summarized in Table~\ref{table.obs}, and total about 216~ks (pn), 277~ks (MOS1), and 289~ks (MOS2). By default, we included events with pattern$\le$4, which are well calibrated. We also used ESAS tools to estimate the quiescent particle background and contamination from solar-wind charge exchange (SWCX), which are spatially uniform and can be subtracted from each image. After removing point sources (see below), for the purposes of image analysis we combined the MOS and pn images to make a single, exposure-corrected image for a soft ($0.4-1.4$~keV) and hard ($2-5$~keV) band, scaled to the MOS2 count rate, using the ESAS {\tt comb} procedure and subtracting the particle background and SWCX emission. The 1.4\,keV cutoff in the soft band was chosen because of the strong instrumental Si and Al lines at 1.4 and 1.75\,keV, and because diffuse hot gas with $kT<1$\,keV primarily emits below 1\,keV.
 
The \Chandra{} data were prepared using the \Chandra{} Interactive Analysis of Observations (CIAO v.4.8). Most of the data reduction is accomplished in the script {\tt chandra\_repro}, which creates a level=2 event file from the primary \Chandra{} data using the standard event patterns (grades). We also used the {\tt deflare} procedure to identify periods of particle flaring, which are ignored for the diffuse analysis. This left a total of 140.7~ks of good time (Table~\ref{table.obs}). We then used the {\tt wavdetect} procedure to identify point sources, using the default 5 scales, and extracted source-free images and a source list for use with the lower resolution \XMM{} data. 
The 0.3-10~keV point-source sensitivity for the combined ACIS data is $F_X \approx 2\times 10^{-16}$~erg~s$^{-1}$~cm$^{-2}$ (assuming a power law spectrum with $\Gamma=2$). 

We detected and masked point sources in the \XMM{} images using both the SAS {\tt edetect\_chain} script and the list of \Chandra{} sources. We masked these sources to a radius corresponding to a fixed level of residual contamination rather than the encircled-energy radius (because a faint source has the same encircled-energy radius as a bright one). We created source masks for sources detected by \Chandra{} within about one arcminute of the disk. 

\subsection{Spectral Modeling Procedure}

For spectral analysis, we defined regions of interest and used the ESAS {\tt mos-spectra} and {\tt pn-spectra} tools to extract source spectra and create spectral response files from each observation after masking the point sources. We used the same regions to extract spectra from the two \Chandra{} data sets with the CIAO {\tt specextract} script. These spectra were then combined using the CIAO {\tt combine\_spectra} script, which also produces the appropriate spectral response files. It is not appropriate to combine the \XMM{} spectra.

We used \textit{Xspec} v12.9.1m \citep{arnaud96} to jointly fit models to the \Chandra{} and \XMM{} spectra. For any given region, this amounts to 19 spectra (six \XMM{} observations with three cameras each, plus the combined ACIS-S spectrum). We restricted fitting to the $0.3-5$\,keV bandpass for the ACIS and MOS spectra, and to $0.4-5$\,keV for the pn, since the pn data below 0.4\,keV may be unreliable. In practice, there is little source emission above 5\,keV, so these bins do not help constrain the parameters of interest. We fitted unbinned spectra using the Cash statistic \citep[C-stat;][]{cash79} and compared them with the \textit{Xspec} {\tt goodness} routine. Since the spectra have high $S/N$, $\chi^2$ gives consistent best-fit parameters, and we report the $\chi^2$ values as a measure of goodness-of-fit for spectra binned to 20~counts per bin. Confidence intervals for each parameter of interest were determined by marginalizing over posterior distributions generated by the \textit{Xspec} {\tt chain} procedure, which implements a Markov-chain Monte Carlo algorithm. The proposal distribution was a Cauchy distribution centered on the initial best-fit parameters, and we ran chains 30,000 steps long (after a 5,000-step burn-in period). We then rescaled the covariance matrix by 0.125, as recommended by the \textit{Xspec} documentation, and repeated the chains. We replaced or extended chains when the heuristic estimates, such as the frequency of repeated values, indicated that a chain was too short. 

Previous works clearly showed that NGC~891 is surrounded by hot gas, so we start from a simple absorbed, isothermal model, adding complexity as needed. Our default photoelectric absorption model is the {\sc phabs} model, our default Solar abundance table is from \citet{grevesse98}, and our default thermal plasma emission model is {\sc apec} v3.0.7 \citep{foster12}. \XMM{} observations can be affected by residual proton flares (flares too weak to be detected as 3$\sigma$ variations in the light curve), so we followed the ESAS procedure of fitting a broken power-law model using a uniform spectral response. This component varies between observations, and is weakly concentrated towards the center of the chip \citep{gastaldello17}. Finally, to allow for small differences in the gain calibration and in the footprints of the spectral extraction regions, we include a constant multiplier for each instrument (with the MOS1 value fixed at 1.0). The value typically ranges from 0.95-1.05 in the pn.

\subsection{Spectral Background Model}

Since the \XMM{} background has vignetted and un-vignetted components, on-field background subtraction for extended sources is inappropriate. Thus, we also defined background regions (about 8\,arcmin off-axis, or $\sim$25\,kpc) and extracted spectra from these regions for each observation for fitting to determine the astrophysical background surface brightness. The surface brightness profile indicates that any halo emission is well below the background in this region and would not significantly affect the background fits. We also extracted a spectrum from about the same regions in the \Chandra{} data (the instrument footprints do not completely overlap), as well as a \ROSAT{} spectrum from 0.25-1~degree around the galaxy using the HEASARC X-ray background tool\footnote{https://heasarc.gsfc.nasa.gov/cgi-bin/Tools/xraybg/xraybg.pl}.

We fitted the background spectra including instrumental components, as descibed in the ESAS documentation, using the following model for the astrophysical soft X-ray background: {\sc apec+phabs(apec+apec+pow)}. These components represent, respectively, the Local Hot Bubble ($kT\equiv 0.1$\,keV, $Z\equiv Z_{\odot}$), the Galactic hot halo ($kT_1 \equiv 0.1$\,keV, $kT_2$ is a free parameter, and $Z/Z_{\odot} \equiv 0.3$), and the unresolved cosmic X-ray background (CXB; $\Gamma \equiv 1.46$). The absorption is fixed at the Galactic value. This model fits the background well, with a reduced $\chi^2=1.02$ for 6952 degrees of freedom. We verified that we obtained consistent parameters when not using \ROSAT{} data. The temperature for the Galactic hot halo is $kT=0.25\pm0.02$\,keV, which is consistent with \citet{henley13}. We include a scaled background as a fixed model in fitting the source spectra, assuming that the astrophysical background has a constant surface brightness in the field of view, except that we allow the \ion{H}{1} halo of NGC~891 to absorb the CXB where appropriate. The instrumental background lines do vary across the chip, so we fit these individually in each source spectrum. Likewise, we included Gaussian lines with zero width to account for SWCX at 0.46, 0.57, 0.65, 0.81, and 0.92~keV, which are \ion{C}{6}, \ion{O}{7}, \ion{O}{8}, \ion{O}{8}, and \ion{Ne}{9} lines. We allow these to vary between observations but not between instruments, then scale the flux appropriately to the source regions. We also scaled the best-fit values for the residual proton flares to account for vignetting and the source region areas \citep{gastaldello17,fioretti16}. 

\section{Image Analysis}
\label{section.images}

The combined \XMM{} images are shown in Figure~\ref{figure.xmm_images} for the soft and hard bands (top and bottom rows). In the left column, we have removed the point sources, smoothed the images with a Gaussian kernel of 3~pixels, clipped the image at 1$\sigma$ above background, and added contours at 2, 4, 8, and 12$\sigma$. To improve the signal of fainter features without creating artifacts, we adaptively binned the images (central column) to $S/N=12$ per bin, using the \citet{sanders16} contour-binning package. We set the contour-following parameter to the minimum to avoid creating artificial structure. The 3$\sigma$ contour is shown as a thick black line in these images. The top-right panel shows the X-ray contours on the 2MASS $H$-band image, as well as H{\sc i} contours. The bottom-right panel show the 3$\sigma$ contours from the soft and hard adaptively binned images. Since the major axis is approximately oriented along the north-south axis, we define north, south, east, and west based on the sky position.

\subsection{Thermal Emission}

The soft image clearly shows the north-south asymmetry in the intensity within 0.5\,arcmin of the disk, which was reported by previous authors \citep{bregman94} and is connected to the higher SFR on the northern side. At lower intensities the asymmetry is much less pronounced (see the 3$\sigma$ contour in Figure~\ref{figure.xmm_images}), which suggests that there is a background halo that is less affected by star formation. There is also an east-west asymmetry, which exists at all galactocentric radii but is most pronounced over the galactic center, where the 3$\sigma$ contour extends to 8\,kpc and the 2$\sigma$ contour extends to 15\,kpc. This asymmetry was noted in the \Chandra{} data at lower heights and lower significance by \citet{li13a}. We tentatively identify this as a one-sided filament, which may be connected to a galactic outflow, based on one-sided extensions seen in starburst galaxies \citep{strickland04,tullmann06}. The decrease in intensity at the midplane reported previously \citep{bregman94} is very pronounced in this image and clearly follows the disk. It is more visible on the north side because of the bright emission just outside the disk. 

The highest intensity comes from the center, but the diffuse emission in the galactic center is dominated by scattered light in the psf wings of an ultraluminous X-ray source (ULX) reported in \citet{hk12}. The large mask in Figure~\ref{figure.xmm_images} extends to the 95\% encircled-energy radius (averaged between the EPIC detectors), which is slightly more than 1~arcmin in radius. However, the ULX was not ``on'' during the deep \Chandra{} exposures, which concentrate the light much better in any case. The \Chandra{} image \citep{strickland04} shows filamentary structure near the disk, and especially above the galactic center, where the brightest filaments make an X-like structure. This supports the hypothesis that the gas seen with \XMM{}, at larger heights, is part of an outflow. There is also bright, extraplanar emission above the star-forming region to the north.

\subsection{Diffuse Hard Excess}

\begin{figure*}
\begin{center}
\includegraphics[width=0.65\textwidth]{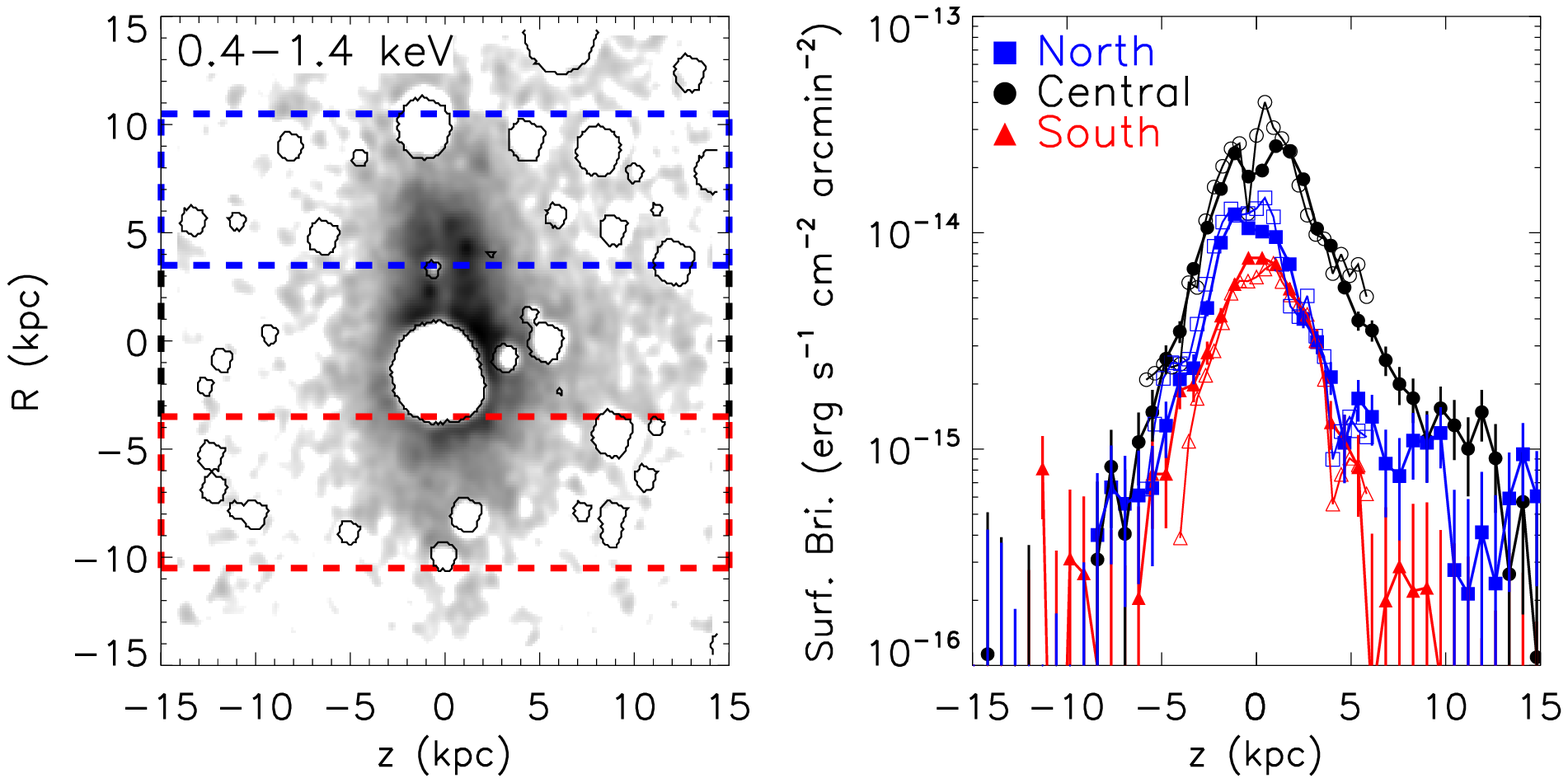}
\includegraphics[width=0.65\textwidth]{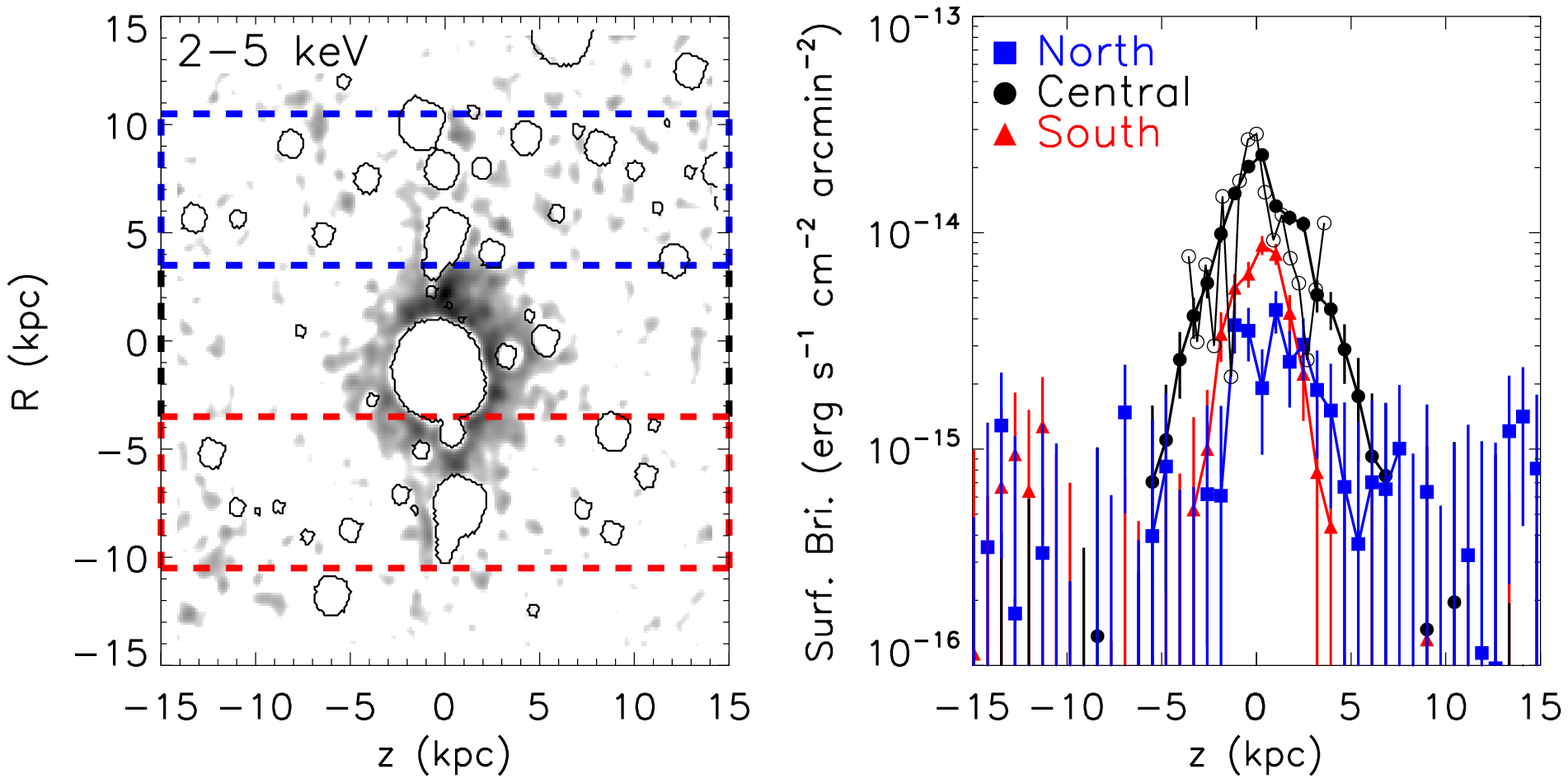}
\caption{Vertical intensity profiles above the midplane in north, central, and south regions (boxes are $15\times 7$~kpc). The top panels show the soft band image and profiles in each zone, while the bottom panels show the hard band. The profiles represent the mean intensity at each distance $z$ from the midplane, and have been rebinned for visual clarity. Filled points are measured from the \XMM{} image, while open points are from \Chandra{}. In the top panel, the \Chandra{} points do not show error bars for visual clarity. The soft emission extends farther to the west (positive $z$) than to the east, and the hard emission is wider around the galactic center than to the north and south.}
\label{figure.height_profile}
\end{center}
\end{figure*}

The 2-5~keV image has lower signal, but there is diffuse hard X-ray emission above the optical disk, with a concentration near the galactic center and an extent of at least 5-6~kpc on each side (Figure~\ref{figure.xmm_images}). The background-subtracted diffuse count rate within $r<2^{\prime}$ of the galactic center is $S_{2-5} = 4.5\pm0.3 \times 10^{-3}$~counts~s$^{-1}$, where the count rate is scaled to the MOS2 detector using the ESAS software. The background subtraction includes removal of the quiescent particle background and CXB, and the source flux excludes masked point sources. This amounts to $F_X = 6.0\times 10^{-14}$~erg~s$^{-1}$~cm$^{-2}$, assuming a power law with $\Gamma=2$, which we assume for the rest of this section.

At least some of this light must be scattered light in the wings of the psf from detected sources, but a comparison between \XMM{} and \Chandra{} indicates that about half of the diffuse flux detected in \XMM{} is astrophysical. For \XMM{}, we estimated the contribution of scattered light by using the aperture-corrected fluxes from all point sources detected within 3~arcmin of the galactic center along with the measured on-axis EPIC encircled-energy profiles to determine the amount of residual light outside each source's mask. The psf shape differs between the detectors, but the fraction of light well outside of the core is about the same. The average residual flux for the combined image is $F_X \approx 3\times 10^{-14}$~erg~s$^{-1}$~cm$^{-2}$. This leads to a MOS2 count rate of $S_{\text{contam}} = 2.2\times 10^{-3}$~counts~s$^{-1}$. The ULX is the largest contributor, and the contamination from the ULX alone is about 17\% of the source flux on average, with 30\% and 12\% for 2011 and 2017, respectively. This does not account for pile-up, which would lead to an underestimate of the scattered light flux. However, the ULX count rates for all epochs are in the range where pile-up is a marginal effect, as verified by an examination of the pattern=0 events with {\tt epatplot}. Thus, the corrected 2-5~keV source flux in the central 2~arcmin is $F_X = (3.1\pm0.2) \times 10^{-14}$~erg~s$^{-1}$.

\Chandra{} has a much tighter psf ($r<0.5^{\prime\prime}$) than \XMM{}, so the scattered light remaining after using the \XMM{} point source masks is negligible. The diffuse 2-5~keV flux from the central 2~arcmin is $S_{2-5} = 2.5\pm0.1 \times 10^{-3}$~counts~s$^{-1}$, or $F_X = (3.2\pm0.1) \times 10^{-14}$~erg~s$^{-1}$, after accounting for the spatially variable background and exposure correction. 
This agrees with the contamination-corrected \XMM{} measurement, and also strongly disfavors backgrounds unique to each instrument as the source. Variation in the CXB is also ruled out, as \Chandra{} resolves more of the CXB than the \XMM{} images. We then used the much smaller \Chandra{} 95\% encircled-energy masks and measured a total 2-5~keV flux of $F_X = (7.6\pm0.1)\times 10^{-14}$~erg~s$^{-1}$~cm$^{-2}$ in the same region. At the distance of NGC~891, this is a 2-5~keV $L_X \approx 10^{39}$~erg~s$^{-1}$. 

This light cannot be unresolved emission from active stars and compact objects. The potential sources of nonthermal X-ray emission include active stars, cataclysmic variables (CVs), high-mass X-ray binaries (HMXBs), and low-mass X-ray binaries (LMXBs). HMXBs are confined to regions of active star formation, and active stars, CVs, and LMXBs follow the distribution of the starlight. Thus, we do not expect a significant extraplanar contribution. Moreover, the \Chandra{} sensitivity is about $L_X > 2.4\times 10^{36}$~erg~s$^{-1}$ at the distance of NGC~891, so the brightest sources (which contribute a large fraction of the flux) would be resolved and masked. 

To estimate the possible contribution we adopted the \citet{lehmer10} relations for the characteristic X-ray binary luminosities: $L_{\text{HMXB}} = 1.62\times 10^{39} \times \text{SFR}$~erg~s$^{-1}$ and $L_{\text{LMXB}} = 9.05\times 10^{28} \times (M_*/M_{\odot}$~erg~s$^{-1}$. For NGC~891, we expect $L_{\text{HMXB}} \approx 6\times 10^{39}$~erg~s$^{-1}$ and $L_{\text{LMXB}} \approx 2\times 10^{39}$~erg~s$^{-1}$. As a late-type galaxy, the contribution from CVs and active stars will be lower \citep{lehmer10}, and NGC~891 does not have a very prominent bulge. X-ray binaries follow a characteristic luminosity function \citep{gilfanov04,mineo12}, so we use this to estimate the amount of undetected X-ray luminosity at the \Chandra{} sensitivity. For the LMXBs this is 0.3-10~keV $L_X \sim 3\times 10^{36}$~erg~s$^{-1}$, and for the HMXBs it is $L_X \sim 10^{36}$~erg~s$^{-1}$. Bearing in mind that the unresolved HMXBs follow star formation and LMXBs the distribution of the starlight, the expected extraplanar contribution is perhaps a few percent of this. This rules out unresolved binaries as the origin of the diffuse excess. Instead, it could be inverse-Compton scattering from cosmic rays, thermal emission from very hot gas, or something else. We return to its origin in Section~\ref{section.discussion}.

\subsection{Minor-axis Profiles}

\begin{deluxetable}{c|cc|cc}
\tablenum{2}
\label{table.scaleheights}
\tabletypesize{\scriptsize}
\tablecaption{Scale Heights from Combined XMM Image}
\tablewidth{0pt}
\tablehead{
\colhead{Region} & \multicolumn{2}{c}{Soft} & \multicolumn{2}{c}{Hard} \\
\colhead{} & \colhead{East} & \colhead{West} & \colhead{East} & \colhead{West}\\
\colhead{} & \colhead{(kpc)} & \colhead{(kpc)} & \colhead{(kpc)} & \colhead{(kpc)}
}
\startdata
North   & $1.8\pm0.3$ & $2.4\pm0.4$ & $0.5\pm0.4$ & $1.1\pm0.8$ \\ 
Central & $1.8\pm0.3$ & $2.9\pm0.4$ & $1.9\pm0.2$ & $2.2\pm0.3$ \\
South  & $1.4\pm0.2$ & $1.6\pm0.2$ & $0.8\pm0.2$ & $0.6\pm0.2$ 
\enddata
\tablecomments{``East'' and ``West'' are approximate and correspond to the left and right side of the midplane in Figure~\ref{figure.height_profile}. Scale heights may be affected by point source masks near the disk.}
\end{deluxetable}

To examine the vertical extent of the emission and the significance of the east-west asymmetry in the soft band, we measured the average vertical profile along the minor axis ($z$) in three sections from north to south (Figure~\ref{figure.height_profile}). The emission is clearly visible to a larger height on the west side of the galaxy than on the east; in the central region, the maximum vertical extent above background is about 8\,kpc on the east and 15\,kpc on the west. This is the candidate filament, which is brighter as well as more extended to the west. There is also more emission in the northeast part of the galaxy than the southeast. We investigated the background in the \XMM{} images, but found no systematic increase towards that region that can explain the excess. Furthermore, the asymmetry persists in the same way in the 2011 image, the combined 2017 exposures, and (to a lesser extent) in the \Chandra{} image, so we conclude that it is real. Figure~\ref{figure.height_profile} also shows the vertical profile for the hard band, where the diffuse emission clearly extends above the disk in the central region but not in the northern or southern regions. 

We measured scale heights in each section by fitting an exponential profile plus constant background, but we caution that the large point source masks likely affect the values. For the soft emission, we ignored the central $|z|<1$~kpc, where the absorption is strong. We also allowed the scale height to differ on the east and west sides, and measured it in the northern, central, and southern regions. The results are given in Table~\ref{table.scaleheights}. There is clearly an east/west asymmetry, which is strongest in the central region. The overall lower scale height in the south occurs because there is less X-ray emission near the disk there, and there is also less star formation in this region. If the emission comes from hot gas, then the scale height of the intensity reflects the square of the density, so the density scale height is about twice as large. In the hard image, the many point source masks suggest that the actual scale heights are known to worse precision than suggested by the error bars quoted here. Nonetheless, the measured scale heights reinforce the conclusion that the extraplanar hard X-ray emission occurs primarily above the central region. 

\subsection{Relative Temperature Map}

\begin{figure*}
\begin{center}
\includegraphics[width=1.05\linewidth]{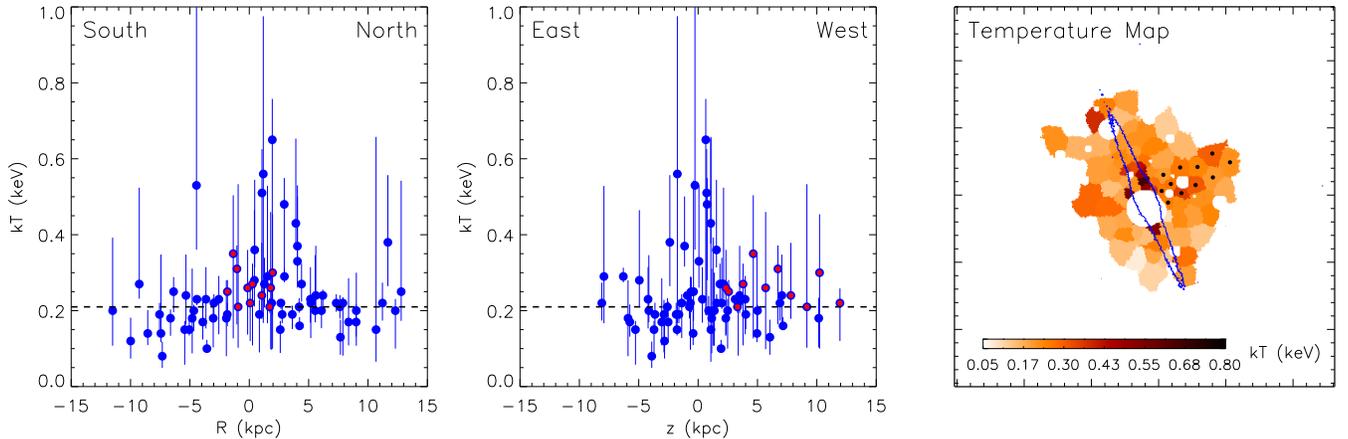}
\caption{Isothermal temperatures measured from adaptive bins with $S/N=25$ and where the average flux is more than 2$\sigma$ above the background. The first two panels show the temperature as a function of distance along the major (left) and minor (center) axes. The red dots indicate temperatures measured from bins in the putative filament region seen in Figure~\ref{figure.xmm_images}, and the dashed line is the mean temperature from all regions. The right panel shows the temperature in each bin as a heat map, with hotter temperatures corresponding to darker colors. The white circles are point source masks, and the filament regions are identified by black dots. The filament region has a higher-than-average temperature for the halo.}
\label{figure.tempmap}
\end{center}
\end{figure*}

We created a relative temperature map by adaptively binning the soft band image to $S/N=25$, extracting 0.4-5\,keV spectra from each bin, and jointly fitting the spectra as described above with an isothermal model. We only fit spectra where the average intensity in the bin was at least 2$\sigma$ above the background. Motivated by the diffuse hard X-rays, we also allowed for the possibility of power-law emission, so the model in each region is {\sc TBabs(apec+pow)}, where the minimum $N_{\text{H}}$ is the Galactic value and the background parameters are fixed. 

The metallicity was fixed at the solar value, which may be incorrect (Section~\ref{section.spectra}), but the temperature is determined by the presence and prominence of various line complexes, and is largely insensitive to the metallicity or normalization, which are degenerate at low $S/N$. An isothermal model may be a poor representation of the temperature structure, but the relative prominence of the \ion{O}{7} (0.57\,keV), \ion{O}{8} (0.65\,keV), \ion{Fe}{17} (0.73, 0.82\,keV) lines, along with the cluster of Fe-L lines at 0.8-1.0\,keV, is a reliable indicator of relative temperature. For example, in simulated pn data from a two-temperature model with matching $S/N$ and $kT_1 = 0.2$~keV and $kT_2 = 1.0$~keV, with the normalization of the hotter component 10\% of that for the cool component, an isothermal fit provides an acceptable fit with a best-fit $kT\approx 0.35$~keV. Then, we repeated the simulation for $kT_1 = 0.2$~keV and $kT_2 = 075$~keV, finding a best-fit $kT \approx 0.25$~keV. To some extent, this depends on the adopted Fe/O abundance ratio, which differs in spiral and elliptical galaxies. We therefore also created a map using the solar metallicity, but with the abundance ratios from \citet{li15}. Although the values changed slightly, the pattern of hotter and cooler regions did not. Thus, the temperature map shows which regions are hotter and cooler than average. We present a more rigorous assessment of the temperature and the metallicity in Section~\ref{section.spectra}. 

\begin{figure}
\begin{center}
\includegraphics[width=1.0\linewidth]{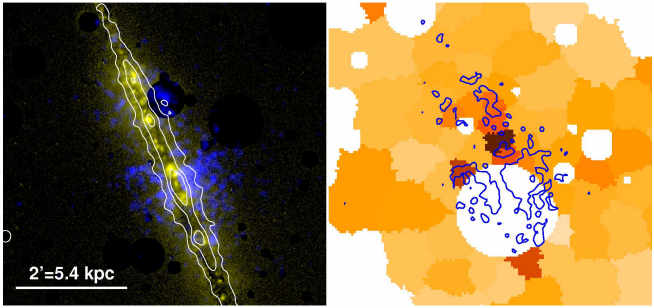}
\caption{\textit{Left}: \Chandra{} 0.3-2\,keV (blue) on the H$\alpha$ (yellow). The \Chandra{} and H$\alpha$ images are clipped at 5$\sigma$ and 3$\sigma$ above background, respectively, to show the brightest emission. 5~GHz contours at 4, 8, and 12$\sigma$ above background are overlaid in white. The X-ray emission is filamentary and clearly concentrated above the bright H$\alpha$ knots, especially in the Galactic center. However, the diffuse ionized gas above the plane is not spatially correlated with the filaments. \textit{Right}: The \Chandra{} filaments (contours in blue are 8$\sigma$ above background) coincide with the hottest (darker) regions in the \XMM{} temperature map (Figure~\ref{figure.tempmap}), indicating that this gas is distinct from the larger halo.}
\label{figure.chandra_filaments}
\end{center}
\end{figure}

\begin{figure}
\begin{center}
\includegraphics[width=1.0\linewidth]{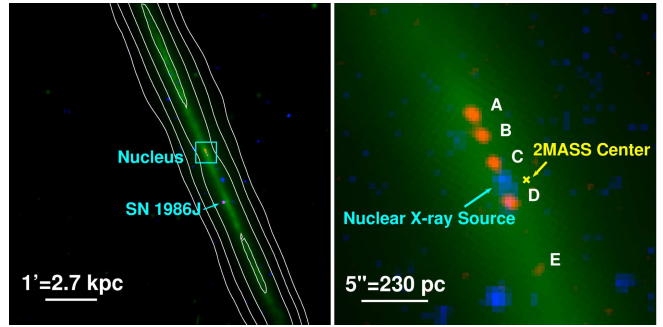}
\caption{An RGB image showing VLA 5~GHz (red), \textit{Spitzer} 5.8~$\mu$m (green), and \textit{Chandra} 0.3-10\,keV point sources (blue). \ion{H}{1} contours have been overlaid at 1, 2, 5, and 10$\times 10^{21}$\,cm$^{-2}$ \citep{oosterloo07}. Radio point sources are detected in the nuclear region and to the south (SN~1986J), but not in the north, despite a sensitivity of about 6~$\mu$Jy/bm ($\nu L_{\nu} \approx 4\times 10^{33}$\,erg\,s$^{-1}$) and the ongoing star formation, which is highlighted by the warm dust. The nuclear region, at right, shows that five radio point sources are detected, but that the nuclear X-ray source \citep{she17} is not. A yellow cross indicates the center of the galaxy as determined by 2MASS; it is consistent with the position of the nuclear X-ray source within the error bars.}
\label{figure.nuclear_image}
\end{center}
\end{figure}

The best-fit temperatures are shown in Figure~\ref{figure.tempmap}, which shows the temperature as a function of projected galactocentric radius $R$, height above the midplane $z$, and as a 2D map using the adaptive bins. The mean temperature measured above 3\,kpc is shown as a dashed line in the first two panels. The temperature rises towards the galactic center, and this is concentrated north of the ULX mask where the soft-band intensity is high. The mean \textit{isothermal} temperature is about 0.21~keV. Figure~\ref{figure.chandra_filaments} shows the brightest \Chandra{} emission relative to the temperature map, as well as the 5~GHz data from our Karl~G.~Jansky Very Large Array (VLA) 2011B-232 and 2012B-165 programs, and archival H$\alpha$ data. The X-ray filaments are clearly associated with nuclear radio emission (see below), H$\alpha$ knots in the disk (although not necessarily with the diffuse ionized gas above the plane), and the hotter regions in the temperature map. This association suggests that the hotter gas is breaking out of the disk into an ambient, cooler halo.

The potential filament to the west is also slightly hotter than the rest of the corona. The average temperature in regions with $z>2$\,kpc and $|R|<2.5$\,kpc is $kT=0.26$\,keV (these regions are highlighted in Figure~\ref{figure.tempmap} with black dots in the map and red dots in the projected profiles). Although the uncertainties for individual regions are high, the values all exceed the mean. This suggests that the filament is a coherent structure, and it may be connected with the brightest X-ray filaments near the center (Figure~\ref{figure.chandra_filaments}).

If we ignore regions within or overlapping the optical disk, so as to isolate the halo emission, there is no significant correlation with either $z$ or $R$ (non-significant Spearman's rank correlation coefficients of $\rho = 0.15$ and 0.12, respectively). Thus, observers within NGC~891 would see little change in the temperature of the corona across the sky, but would measure a strong gradient in the emission measure. This is very similar to the situation for the Milky Way \citep{henley13,miller15}. 

\subsection{Galactic Nucleus}

\citet{she17} reported a nuclear \Chandra{} X-ray source as a candidate low-luminosity AGN. We examined high resolution 5~GHz VLA data from our 2012B-165 program to search for a radio counterpart. We did not find one, down to a limiting sensitivity of 5.5$\mu$Jy~beam$^{-1}$, but found several other point sources in the nucleus (which are labeled A--E in Figure~\ref{figure.nuclear_image}). These sources are exactly on the midplane and so there is no soft X-ray emission (due to absorption by the disk), but we find a potential X-ray counterpart for the radio source directly to the south of the nuclear X-ray source. The radio luminosities of sources A--E are $\nu L_{\nu} = 1.9, 2.0, 2.6, 1.7$, and $0.41\times 10^{35}$~erg~s$^{-1}$, respectively.  The rather faint Source~E is detected at $6.2\sigma$. These are probably supernova remnants, similar to (but dimmer than) SN~1986J to the south. There is no data at comparable resolution at any other radio frequency, so we cannot measure the spectral indices. 

There is a chance that one or more of the radio sources (or the central X-ray source) is seen in projection and not actually nuclear. However, the wider, diffuse radio emission peaks within this region (Figure~\ref{figure.chandra_filaments}), as does the \textit{Spitzer} 5.8~$\mu$m intensity, so it is more likely that there is rapid star formation at the nucleus. This emission also coincides with a nuclear molecular ring \citep{garcia-burillo92}. This would explain the presence of the \Chandra{} filaments (Figure~\ref{figure.chandra_filaments}), which appear similar to those seen in galaxies with nuclear starbursts \citep{strickland04}. 

\section{Spectral Analysis}
\label{section.spectra}

\begin{figure}
\begin{center}
\includegraphics[width=1\linewidth]{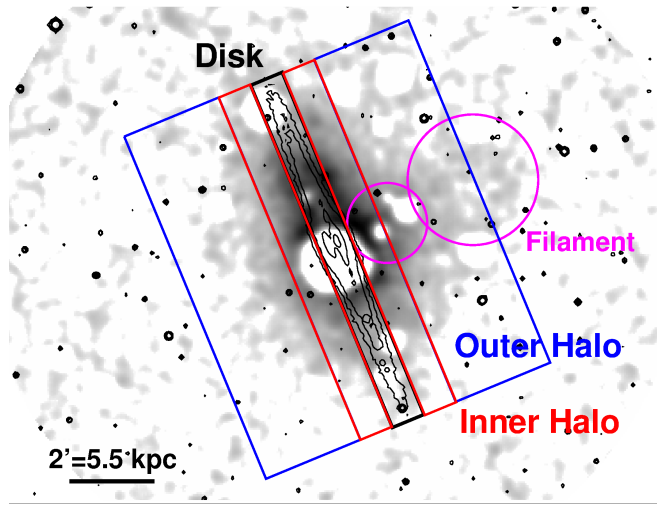}
\caption{Regions used for spectral extraction. For the inner and outer halo regions we combined the spectra from each side of the disk. For the filament regions (magenta circles), we fit the bottom and top separately, as described in the text. The dividing line between the inner and outer halo is about 1.1~kpc above the midplane, above which assuming the Galactic column $N_{\text{H}} = 6.8\times 10^{20}$~cm$^{-2}$ is valid.}
\label{figure.extraction_zone}
\end{center}
\end{figure}

We fitted models to spectra extracted from several regions to measure the temperature and metallicity of the gas in and around the galaxy. These regions are shown in Figure~\ref{figure.extraction_zone}. We defined the halo as the region above 1.1\,kpc from the midplane, and subdivided it into the inner halo ($1.1 < z < 3.3$\,kpc) and outer halo ($3.3 < z < 10$\,kpc), with the boundary set where the intrinsic $N_{\text{H}} \approx 5\times 10^{19}$\,cm$^{-2}$, which is less than 10\% of the Galactic $N_{\text{H}} = 6.8\times 10^{20}$\,cm$^{-2}$ \citep{kalberla05}. Thus, in the outer halo we fix the absorbing column at Galactic, whereas inside it is a free parameter. We determined the best-fit models for each zone separately, then jointly modeled both zones with $N_{\text{H}}$ as a free parameter for the inner region. Since thousands of counts are needed for precise metallicities, we restrict the analysis of the metallicity to the inner and outer halo regions. In the remainder of this section, we describe our approach to fitting the spectra, present our results, and discuss possible sources of error. The results are summarized in Table~\ref{table.spectra}.

\subsection{Halo Spectra}

\begin{figure}
\begin{center}
\includegraphics[width=1\linewidth]{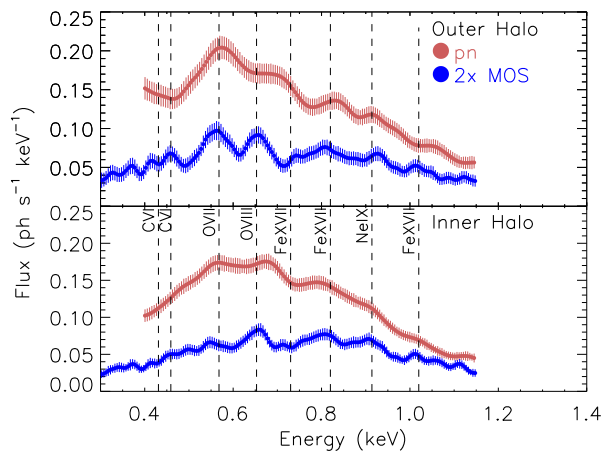}
\caption{Combined \XMM{} spectra for the outer halo ($3.3<z<10$\,kpc) and inner halo ($1.1<z<3.3$\,kpc) after subtracting the best-fit background model, scaled to each region. These ``folded'' spectra include the energy dependence of the effective area. The MOS 1+2 spectra were combined into a single spectrum and scaled to better match the pn, and each spectrum has been lightly smoothed for visual clarity. The emission lines used to determine the temperature and metallicity are marked.}
\label{figure.halo_spectra}
\end{center}
\end{figure}

Figure~\ref{figure.halo_spectra} shows the combined, folded $0.3-1.2$~keV MOS (and $0.4-1.2$~keV pn) spectra for all epochs, with the best-fit Galactic and instrumental backgrounds subtracted. The spectra have been lightly smoothed for visual clarity. The diagnostic emission-line complexes include \ion{O}{7} (0.57\,keV), \ion{O}{8} (0.65\,keV), \ion{Fe}{17} (0.82, 1.02\,keV), and \ion{Ne}{9} (0.92\,keV), which unambiguously indicate a thermal plasma. The prominence of the \ion{O}{7} and \ion{O}{8} lines, especially in the outer halo, indicates a temperature of $kT \lesssim 0.25$~keV. The spectra from the inner halo have a larger proportion of flux at higher energies, which could either mean hotter gas or more photoelectric absorption (which predominantly affects softer photons). Finally, the continuum above 1\,keV is too high for a plasma with $kT < 0.3$\,keV and there are no strong lines from \ion{Fe}{21}-\ion{Fe}{25} that would indicate a hotter plasma, so we conclude that there is a non-thermal component. This is in agreement with the image analysis in Section~\ref{section.images}.

\begin{figure}
\begin{center}
\includegraphics[width=1\linewidth]{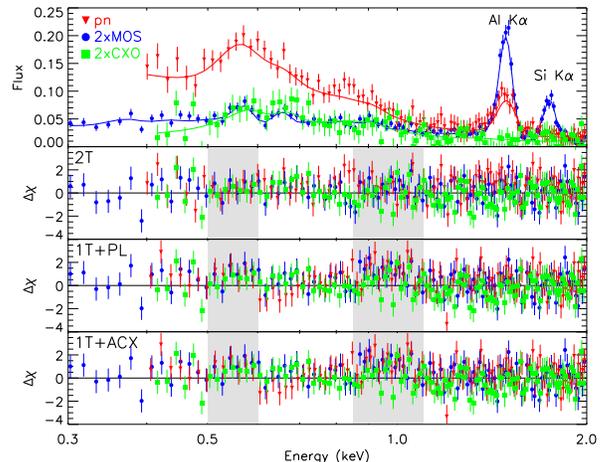}
\caption{\textit{Top}: pn, MOS, and ACIS data (red, blue, and green) from the outer halo are shown with the best-fit 2T model overlaid. The data have been combined and rebinned for visual clarity. The strong instrumental Al and Si lines are marked. \textit{Bottom}: Residuals for the 2T model, 1T+PL model, and 1T+ACX model. The shaded regions show where the 2T model is an improvement over the 1T models.}
\label{figure.resid}
\end{center}
\end{figure}

Based on these features, we began by modeling the spectrum as the sum of an isothermal (1T) and power-law model ({\sc phabs(apec+phabs(pow))}). We tried variants where the power-law index was fixed at the best-fit value as measured in the $E=2-5$~keV bandpass and where it is free to vary, and we obtained better fits when it was allowed to vary. The best-fit thermal parameters were $kT = 0.23\pm0.01$~keV and $Z/Z_{\odot} = 0.15_{-0.04}^{+0.12}$. However, the 1T+PL model is ruled out using the criterion of $p<0.05$ for the $\chi^2$ value, and it leaves positive residuals near 0.5~keV and 1~keV that indicate that the model does not fit the emission lines well (Figure~\ref{figure.resid}). 

\begin{figure*}
\begin{center}
\includegraphics[width=1\linewidth]{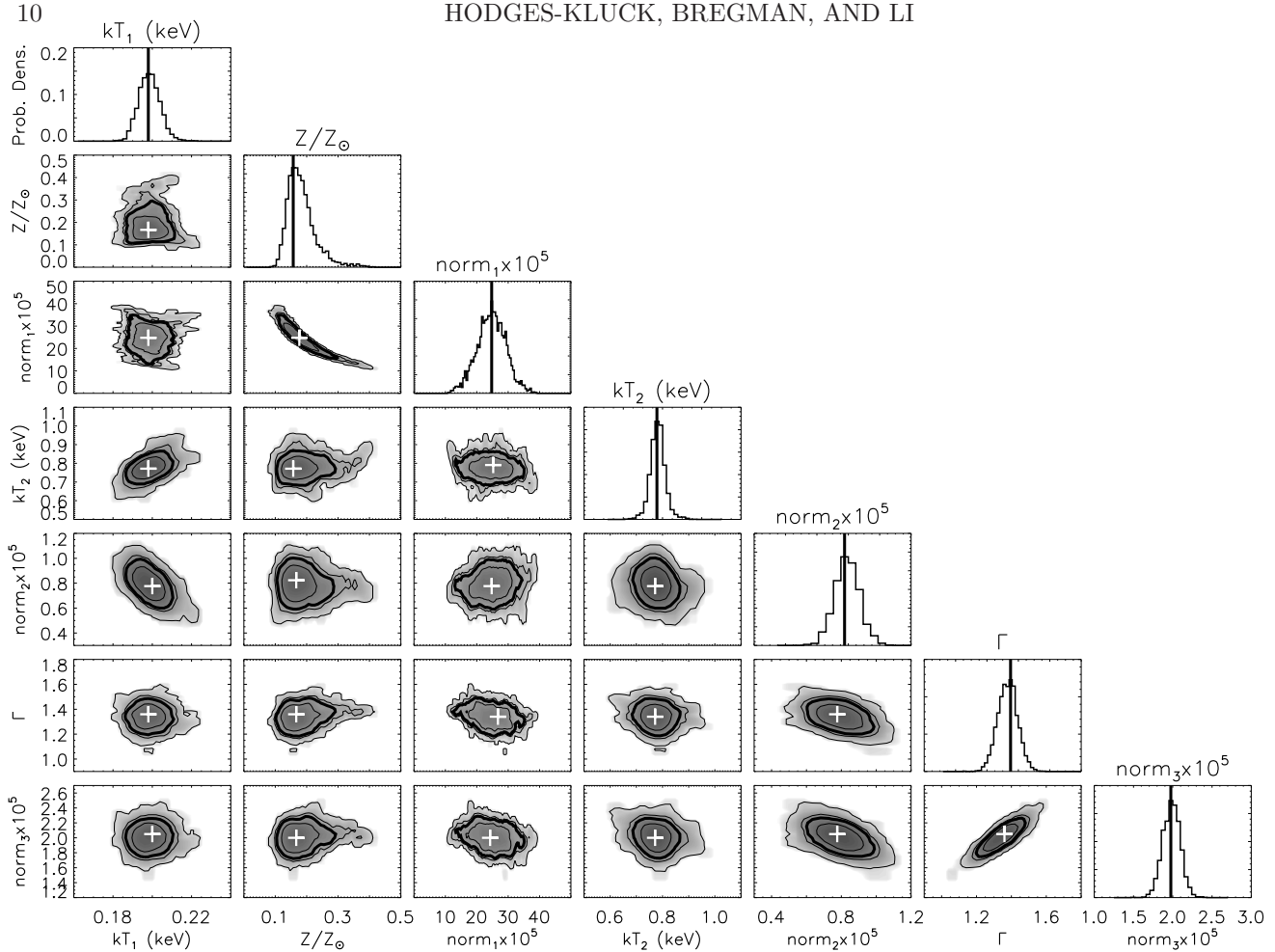}
\caption{Posterior distributions for model parameters in the fit to the outer halo. The thin contours are the 1, 2, and 3$\sigma$ (68\%, 95\%, and 99.74\%) credible regions, while the thick contour is the 90\% credible region. White crosses and the black vertical lines indicate the most likely values for each projection.}
\label{figure.halo1_mcmc}
\end{center}
\end{figure*}

Thus, we added a second thermal component  (2T, {\sc phabs(apec+apec+phabs(pow))}). This produces an acceptable fit in each halo region and leaves flatter residuals. A summary of the model parameters for the 2T+PL models we tried is given in Table~\ref{table.spectra}. Figure~\ref{figure.halo1_mcmc} shows the MCMC sampling and best-fit parameters for the 2T, PL-free model in the outer halo. In both halo regions, $kT_1 = 0.20\pm0.01$~keV, and $Z/Z_{\odot} \approx 0.14$. $kT_2 = 0.77$~keV in the outer halo and $0.68$~keV in the inner halo, and the metallicity cannot be independently determined. This is because the hotter component contributes only about 10\% as much flux as the cooler component, and so all the emission is from lines (the continuum is invisible). We fixed the metallicity of this component at Solar for this reason, but we note that if it is instead tied to the metallicity of the cooler component only the normalization of the hotter material changes.

We also tried more complex models, including adding a charge-exchange model to the 2T model and a multi-T (3T or 4T) model. \citet{zhang14} found that much of the X-ray emission from the soft X-ray nebula around M82 arises from charge exchange. Unlike NGC~891, M82 is a starburst galaxy with a strong galactic wind, but the massive \ion{H}{1} halo around NGC~891 implies the presence of many hot-cool interfaces where charge exchange could occur, albeit to a lesser extent because of the lower velocities. We used the ACX~v1.0.4 model from \citet{smith12} ({\sc phabs(apec+acx+phabs(pow))}), with the temperature and metallicity tied to the {\sc apec} component, but this is a worse fit than the 2T model (Figure~\ref{figure.resid}). When we tried multi-T models, they always collapsed to the 2T case; they do not improve the fit. Hence, we conclude that the 2T model is the simplest model that adequately fits the data, and that obvious sources of additional complexity are not favored.

In the 2T+PL model, the best-fit absorbing column in the inner halo is consistent with the absorption expected from NGC~891, based on the \ion{H}{1} map, and the main reason for the difference between $kT_2$ in the inner and outer halo is the pn spectrum in the inner halo; when fitting the spectra from each instrument individually the inner-halo MOS spectrum agrees with both the MOS and pn fits in the outer halo (Table~\ref{table.spectra}). Thus, we conclude that the spectra from both regions can be fitted jointly with a single model, allowing for differences in instrumental backgrounds and absorbing column. 

The combined fit is acceptable and has the following best-fit parameters (Table~\ref{table.spectra}): $kT_1 = 0.20\pm0.01$~keV, $Z/Z_{\odot} = 0.14\pm0.03$, $kT_2 = 0.73\pm0.04$~keV, and $\Gamma=1.4\pm0.1$. We adopt this parameter set as our fiducial model. The measured 0.3-10~keV luminosities are $L_X = 2.9\times 10^{39}$~erg~s$^{-1}$ for the cooler component, $L_X = 5.3\times 10^{38}$~erg~s$^{-1}$ for the hotter component, and $L_X = 3.8\times 10^{39}$~erg~s$^{-1}$ for the nonthermal component. 

We also fitted a 2T {\sc vapec} model to determine the oxygen-like and iron-like abundances separately. We applied this model to both the whole halo fit and the outer halo individually. In the outer halo, we found $kT_1 = 0.20\pm0.01$\,keV, $Z_{\text{O}}/Z_{\text{O},\odot} = 0.17_{-0.06}^{+0.04}$, and $Z_{\text{Fe}}/Z_{\text{Fe},\odot} = 0.21_{-0.08}^{+0.20}$, which are consistent with the single $Z/Z_{\odot}$ (solar abundance ratios). We found similar results for the whole halo, with $kT_1 = 0.20\pm0.01$\,keV, $Z_{\text{O}}/Z_{\text{O},\odot} = 0.14_{-0.02}^{+0.04}$, and $Z_{\text{Fe}}/Z_{\text{Fe},\odot} = 0.17_{-0.03}^{+0.06}$. This is consistent with the slightly lower single $Z/Z_{\odot}$ measured from the fit to the whole halo. 

\subsection{Halo Properties}

We used the fiducial model to derive some basic quantities for the halo, including the mass and pressure of the hot gas. The best-fit parameters and these properties are summarized in Table~\ref{table.halo_properties}. 

First, we corrected for the fraction of the halo region covered by point source masks by scaling the measured fluxes by the ratio of the total-to-unmasked area in the halo region. This leads to 0.3-10 (0.01-10)~keV luminosities of $L_X \approx 3.2\times 10^{39}$ ($1.2\times 10^{40}$)~erg~s$^{-1}$ for the $kT=0.20$~keV component and $L_X \approx 7.5\times 10^{38}$~($9.8\times 10^{38}$)~erg~s$^{-1}$ for the hotter component. The nonthermal component must also be corrected for light scattered into the wings of detected (and masked) point sources. As in Section~\ref{section.images}, we estimated the contamination fraction in each region from the aperture-corrected point source fluxes, the existing source masks, and the EPIC psf curves. This amounts to 15\% for the outer halo and 60\% for the inner halo. When combining this effect with the scaling for the masked area, we obtain a 0.3-10 (0.01-10)~keV $L_X \approx 3.1\times 10^{39}$ ($3.3\times 10^{39}$)~erg\,s$^{-1}$ for the nonthermal component.

We estimated the mass and average density of gas detected by \XMM{} by assuming that the cooler gas fills a cylinder with $R=10$~kpc and a height of $z=9$~kpc on each side of the galaxy, and that the hotter gas fills a cylinder with $R=6$~kpc and $z=4$~kpc, based on where X-ray emission is detected and where star formation occurs. Assuming a mean molecular weight for ionized gas of $\mu = 0.62$, the {\sc apec} normalization $\text{norm}=6.9\times 10^{-4}$ for the whole halo leads to $M \approx 2.3\times 10^8 M_{\odot}$ for the cooler component. This is comparable to previous estimates \citep{bregman94,hk13}. The mean density is $\bar{n} = 2.2\times 10^{-3}$~cm$^{-3}$, which corresponds to a mean $\bar{P} = 7\times 10^{-13}$~dyne~cm$^{-2}$. The cooling rate is estimated as $\dot{M}_{\text{cool}} = L_X/e$, where $e=3kT/2\mu m_{\text{H}}$. The 0.01-10~keV $L_X = 1.5\times 10^{40}$~erg~s$^{-1}$ for the virialized component, leading to $\dot{M}_{\text{cool}} = 0.5 M_{\odot}$~yr$^{-1}$ within $1.1 < |z| < 8$~kpc.

The hotter gas has a {\sc apec} normalization of $1.9\times 10^{-5}$, which corresponds to $M \approx 10^7 M_{\odot}$, $\bar{n} = 9\times 10^{-4}$~cm$^{-3}$, and $\bar{P} = 10^{-12}$~dyne~cm$^{-2}$. These values are insensitive to the gross geometry, such as an exponential disk or $\beta$-model \citep{cavaliere76}, since the volume in which X-ray emission is detected is fixed. However, they are sensitive to the filling factor. We expect the virialized component to be nearly volume-filling, but the hotter gas may not be as it is connected to star formation in the disk. 

50\% of the cooler gas and 40\% of the hotter gas comes from the outer halo, while this region covers 75\% of the projected area between 1.1-10~kpc (not accounting for point source masks, which cover 16\% of the inner halo and 5\% of the outer halo). The hotter gas is concentrated near the disk, as shown in Figure~\ref{figure.tempmap}.

\begin{turnpage}
\begin{deluxetable*}{lcccccccccccc}
\tablenum{3}
\tabletypesize{\scriptsize}
\tablecaption{Halo Spectra 2T Model Fits}
\tablewidth{0pt}
\tablehead{
\colhead{Model} & \colhead{$N_{\text{H}}$} & \colhead{$kT_1$} & \colhead{$Z_1$} & \colhead{norm$_1$} & 
\colhead{$kT_2$} & \colhead{$Z_2$} & \colhead{norm$_2$} & \colhead{$N_{\text{H,pow}}$} & 
\colhead{$\Gamma$} & \colhead{PL norm} &  \colhead{$\chi^2$} & \colhead{dof} \\
\colhead{} & \colhead{($10^{22}$ cm$^{-2}$)} & \colhead{(keV)} & \colhead{($Z/Z_{\odot}$)} & \colhead{($10^{-5}$)} & \colhead{(keV)} & 
\colhead{($Z/Z_{\odot}$)} & \colhead{($10^{-5}$)} & \colhead{($10^{22}$ cm$^{-2}$)} & \colhead{} & 
\colhead{($10^{-5}$)} & \colhead{} & \colhead{} \\
\colhead{(1)} &  \colhead{(2)} &  \colhead{(3)} &  \colhead{(4)} &  \colhead{(5)} &  \colhead{(6)} &  \colhead{(7)} &  
\colhead{(8)} &  \colhead{(9)} &  \colhead{(10)} &  \colhead{(11)} &  \colhead{(12)} &  \colhead{(13)} 
}
\startdata
\multicolumn{13}{c}{\sc Outer Halo}\\
\cline{1-13}\\
PL Fixed	& 0.068 (f)	 & $0.19_{-0.01}^{+0.02}$ & $0.14_{-0.03}^{+0.05}$ & $32_{-8}^{+6}$ & $0.80_{-0.04}^{+0.16}$ & 1 (f) & $1.1\pm0.1$ & $<$0.06 & 1.1 (f) & 1.4 (f) & 3168.2 & 3058 \\
PL Free	& 0.068 (f)	 & $0.20\pm0.01$ & $0.16_{-0.05}^{+0.09}$ & $25\pm8$ & $0.77_{-0.06}^{+0.07}$ & 1 (f) & $0.9_{-0.2}^{+0.1}$ & $<$0.06 & $1.3\pm0.1$ & $1.9\pm0.2$ & 3124.0 & 3056 \\
MOS only	& 0.068 (f)	 & $0.19\pm0.01$ & $0.28_{-0.12}^{+0.44}$ & $19_{-5}^{+1}$ & $0.75\pm0.09$ & 1 (f) & $0.9\pm0.1$ & $<$0.1 & $1.3\pm0.2$ & $1.4\pm0.2$ & 1680.1 & 1574 \\
PN only	& 0.068 (f)	 & $0.20_{-0.01}^{+0.02}$ & $0.13_{-0.04}^{+0.09}$ & $27\pm11$ & $0.76_{-0.09}^{+0.14}$ & 1 (f) & $0.9_{-0.3}^{+0.2}$ & $<$0.1 & $1.2\pm0.2$ & $1.8_{-0.2}^{+0.3}$ & 1291.8 & 1365 \\  
2017 only  & 0.068 (f)	 & $0.20_{-0.01}^{+0.02}$ & $0.11_{-0.03}^{+0.08}$ & $34_{-12}^{+9}$ & $0.8\pm0.1$ & 1 (f) & $0.7\pm0.2$ & $<$0.1 & $1.1_{-0.1}^{+0.2}$ & $1.5\pm0.2$ & 2214.1 & 2098 \\
Pattern=0	& 0.068 (f)	 & $0.20\pm0.01$ & $0.14_{-0.04}^{+0.06}$ & $22_{-5}^{+7}$ & $0.77_{-0.05}^{+0.13}$ & 1 (f) & $0.68_{-0.08}^{+0.11}$ & $<$0.2 & $1.3\pm0.2$ & $1.4\pm0.2$ & 2607.7 & 2596 \\  
MEKAL	& 0.068 (f) & $0.21\pm0.01$ & $0.15_{-0.05}^{+0.04}$ & $19_{-4}^{+10}$ & $0.69_{-0.06}^{+0.09}$ & 1 (f) & $0.55_{-0.12}^{+0.09}$ & $<$0.1 & $1.2\pm0.01$ & $1.8\pm0.2$ & 3224.9 & 3058 \\
VAPEC\tablenotemark{a}	& 0.068 (f) & $0.20\pm0.01$ & $0.17_{-0.06}^{+0.04}$ & $25_{-4}^{+9}$ & $0.8_{-0.1}^{+0.2}$ & 1 (f) & $0.8\pm0.1$ & $<$0.08 & $1.4\pm0.2$ & $2.1\pm0.2$ & 3143.2 & 3054 \\
           	& & 			   & $0.21_{-0.08}^{+0.20}$ &                          &                                &         &                   &               &                   &                     &           & \\
\cline{1-13}\\
\multicolumn{13}{c}{\sc Inner Halo}\\
\cline{1-13}\\
PL Fixed	& $0.14\pm0.04$ & $0.20\pm0.01$ & $0.15_{-0.03}^{+0.07}$ & $30_{-10}^{+20}$ & $0.65_{-0.06}^{+0.04}$ & 1 (f) & $0.81_{-0.06}^{+0.13}$ & $<$0.1 & 1.6 (f) & 2.12 (f) & 1636.3 & 1562 \\
PL Free	& $0.16\pm0.02$ & $0.20\pm0.01$ & $0.11_{-0.03}^{+0.04}$ & $46_{-9}^{+12}$ & $0.68\pm0.04$ & 1 (f) & $0.93_{-0.08}^{+0.11}$ & $<$0.1 & $1.4\pm0.1$ & $1.4\pm0.1$ & 1550.8 & 1560 \\
MOS only	& $0.14\pm0.03$ & $0.21\pm0.02$ & $0.12_{-0.04}^{+0.06}$ & $39_{-9}^{+6}$ & $0.74\pm0.08$ & 1 (f) & $0.7\pm0.1$ & $<$0.05 & $1.6\pm0.1$ & $1.8\pm0.2$ & 776.8 & 733 \\
PN only	& $0.10_{-0.03}^{+0.1}$ & $0.20\pm0.01$ & $0.13_{-0.03}^{+0.06}$ & $27_{-8}^{+5}$ & $0.67\pm0.05$ & 1 (f) & $0.9\pm0.1$ & $<$0.1 & $1.5\pm0.1$ & $1.4\pm0.2$ & 698.8 & 708 \\
\cline{1-13}\\
\multicolumn{13}{c}{\sc Outer+Inner Halo}\\
\cline{1-13}\\
PL fixed 	& $0.12\pm0.02$ & $0.203\pm0.006$ & $0.14\pm0.03$ & $52_{-10}^{+13}$ & $0.72\pm0.05$ & 1 (f) & $1.5\pm0.1$ & $<$0.04 & 1.4 (f) & $3.1\pm0.9$ & 4516.4 & 4439 \\
PL free	& $0.11\pm0.02$ & $0.199\pm0.008$ & $0.14\pm0.03$ & $69\pm6$ & $0.71\pm0.04$ & 1 (f) & $1.9\pm0.1$ & $<$0.06 & $1.4\pm0.1$ & $3.5\pm0.4$ & 4499.9 & 4438 \\
MOS only	& $0.14\pm0.03$ & $0.20\pm0.01$ & $0.15\pm0.04$ & $70\pm10$ & $0.73\pm0.06$ & 1 (f) & $1.6\pm0.3$ & $<$0.07 & $1.5\pm0.1$ & $3.7\pm0.3$ & 2413.56 & 2315\\
PN only	& $0.09_{-0.02}^{+0.1}$ & $0.21\pm0.01$ & $0.12_{-0.02}^{+0.03}$ & $55\pm5$ & $0.76\pm0.07$ & 1 (f) & $1.1\pm0.1$ & $<$0.08 & $1.5\pm0.1$ & $2.8\pm0.2$ & 2000.5 & 2079 \\
VAPEC\tablenotemark{a} & $0.14\pm0.03$ & $0.196\pm0.004$ & $0.13_{-0.02}^{+0.04}$ & $85\pm7$ & $0.74\pm0.04$ & 1 (f) & $1.9\pm0.1$ & $<$0.05 & $1.3\pm0.2$ & $3.2\pm0.2$ & 4506.6 & 4436 \\
                                     &                        &                           & $0.15_{-0.03}^{+0.04}$ &  & & & & & & & & \\
\cline{1-13}\\
\multicolumn{13}{c}{\sc Other Regions}\\
\cline{1-13}\\
Disk 	& $0.3\pm0.1$ & $0.22\pm0.03$ & $0.13_{-0.06}^{+0.08}$ & $4\pm2$ & $0.73\pm0.05$ & 1 (f) & $0.27\pm0.02$ & $<$0.2 & 1.5 (f) & 0.2 (f) & 415.7 & 435 \\
Filament Bot.	& $0.15\pm0.08$ & $0.21\pm0.02$ & $0.14_{-0.05}^{+0.09}$ & $6.7_{-0.8}^{+1.5}$ & $0.63\pm0.09$ & 1 (f) & $0.5_{-0.1}^{+0.3}$ & $<$0.2 & $1.7\pm0.2$ & $0.63\pm0.06$ & 671.2 & 626\\
Filament Top          & 0.068 (f) & $0.21\pm0.02$ & $0.12_{-0.05}^{+0.11}$ & $11_{-5}^{+4}$ & $0.75_{-0.09}^{+0.15}$ & 1 (f) & $0.24\pm0.08$ & $<$0.2 & $1.4\pm0.2$ & $0.17\pm0.07$ & 692.5 & 677 \\
\enddata
\tablenotetext{a}{The first line is $Z_{\text{O}}$ and the second is $Z_{\text{Fe}}$.}
\tablecomments{\label{table.spectra} Best-fit parameters for the various models. Error bars represent the 90\% confidence intervals from MCMC sampling. A ``f'' denotes a frozen parameter. The normalizations for the whole halo are the sum of the inner and outer halo components for those fits, and the $N_{\text{H}}$ value for the whole halo is for the inner component only.}
\end{deluxetable*}
\end{turnpage}

\begin{deluxetable}{lcc}
\tablenum{4}
\tabletypesize{\scriptsize}
\tablecaption{NGC 891 Halo Properties}
\tablewidth{0pt}
\tablehead{
\colhead{Quantity} & \colhead{Units} & \colhead{Value} 
}
\startdata
\multicolumn{3}{c}{Virialized Halo}\\
\hline
$kT_1$ & (keV) & $0.20\pm0.01$\\
$Z/Z_{\odot}$ &  & $0.14\pm0.03$(stat)$^{+0.08}_{-0.02}$(sys) \\
$M$     & ($M_{\odot}$) & $2.3\times 10^8$ \\
$\bar{P}_{\text{th}}$ & (dyne~cm$^{-2}$) & $7\times 10^{-13}$ \\
$L_X (0.3-10)$   & (erg~s$^{-1}$) & $3.2\times 10^{39}$ \\
$L_X (0.01-10)$ & (erg~s$^{-1}$) & $1.5\times 10^{40}$ \\
$\dot{M}_{\text{cool}}$ & ($M_{\odot}$~yr$^{-1}$) & 0.5 \\
$h_{\text{th}}$ & (kpc) & 4-6 \\
\hline
\multicolumn{3}{c}{Hotter Component}\\
\hline
$kT_2$ & (keV) & $0.71\pm0.04$ \\
$Z/Z_{\odot}$ &  & 1 (fixed)\\
$M$     & ($M_{\odot}$) & $10^7$ \\
$\bar{P}_{\text{th}}$ & (dyne~cm$^{-2}$) & $10^{-12}$ \\
$L_X (0.3-10)$  & (erg~s$^{-1}$) & $7.5\times 10^{38}$ \\
$L_X (0.01-10)$ & (erg~s$^{-1}$) & $9.8\times 10^{38}$ \\
\hline
\multicolumn{3}{c}{Nonthermal Emission}\\
\hline
$\Gamma$ & & $1.4\pm0.1$ \\
$L_X (0.3-10)$   & (erg~s$^{-1}$) & $3.1\times 10^{39}$\\
$L_X (0.01-10)$ & (erg~s$^{-1}$) & $3.3\times 10^{39}$
\enddata
\tablecomments{\label{table.halo_properties} Derived values are based on the fiducial model, which is the 2T, PL-free model for the whole halo in Table~\ref{table.spectra}.  All luminosities are corrected for point source masks.
}
\end{deluxetable}

\subsection{Disk and Filament Spectra}

We also extracted spectra from the disk itself and the putative filament to the west (Figure~\ref{figure.extraction_zone}), and fitted them with a 2T model in an analogous approach to the previous section. The filament was fitted in two zones corresponding to the inner and outer halo. The best-fit parameters are listed in Table~\ref{table.spectra}. Remarkably, the temperatures and abundances agree well with the halo spectra, and the $kT_1 = 0.2$~keV, $Z/Z_{\odot} = 0.12$ gas appears to persist into the disk region. It is likely that there is more enriched gas in the disk that cannot be seen because of the strong \ion{H}{1} absorption. The best-fit $N_{\text{H}} = 2\times 10^{21}$~cm$^{-2}$ for the disk region is lower than the total \ion{H}{1} column, so the $kT_1 = 0.2$~keV component is from the near side of the disk, or possibly in front of it. Since unresolved X-ray binaries do not significantly contribute (Section~\ref{section.images}), and almost all emission from the active stars in the disk would be absorbed by the neutral gas in the disk, we do not include these components in our fit. Scattered light in the wings of detected point sources contributes to the measured luminosity of the power law, but does not strongly affect the thermal component. 

\begin{figure}
\begin{center}
\includegraphics[width=1\linewidth]{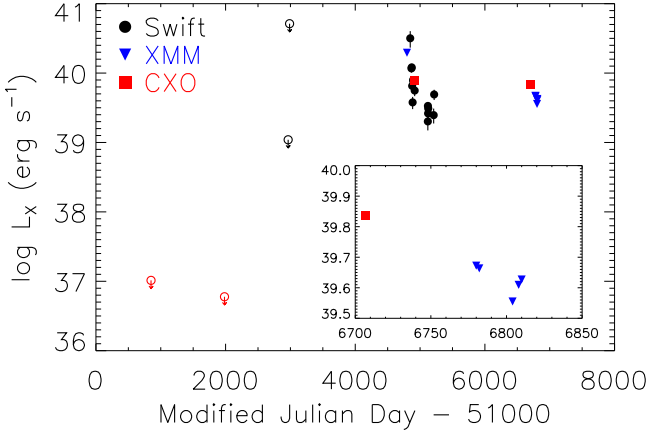}
\caption{The 0.3-10~keV ULX luminosity since July 6, 1998 (MJD 51000). NGC~891 was observed early on in the \Chandra{} mission and no source was detected at the location of the ULX, which we first saw in the 2011 \XMM{} observation. The inset shows the rightmost points, which correspond to a recent \Chandra{} observation (red) and the 2017 \XMM{} data presented here (blue). While the ULX is no longer at $L_X > 10^{40}$~erg~s$^{-1}$, it has not continued the decline seen in the 2012 \textit{Swift} data, or there has been another flare whose peak was missed.}
\label{figure.ulx_fluxes}
\end{center}
\end{figure}

After accounting for point source masks, the bottom section of the filament region covers about 6\% of the total halo area, and it accounts for 10\% of the cooler component, 28\% of the hotter component, and 18\% of the nonthermal component in the fiducial model. Meanwhile, the upper section of the filament covers 19\% of the total halo area and accounts for 16\% of the cooler component, 13\% of the hotter component, and 5\% of the nonthermal component. In total, the filament regions cover 24\% of the halo area and contain about 25\% of the cooler gas and more than 40\% of the hotter gas. Thus, the above-average \textit{isothermal} temperatures in this region seen in Figure~\ref{figure.tempmap} are because a larger fraction of hotter gas in the region instead of a rise in the ambient halo gas temperature. This strengthens the interpretation of the filament as a physical structure, and it is likely that the hot gas that is concentrated in the bottom part of the filament has displaced and compressed the ambient gas, which leads to the increased surface brightness. We return to the topic of its origin in Section~\ref{section.discussion}.

\subsection{ULX}

We fitted the spectrum of the ULX for each epoch, following the approach in \citet{hk12} by using the slim, multi-colored accretion disk model in which $T(r) \propto r^{-p}$, where $p=0.75$ is the standard multi-colored disk \citep[{\sc TBabs(diskpbb)};][]{mineshige94}. The free parameters are the absorbing column, the temperature at the innermost edge of the accretion disk ($T_{\text{in}}$), and the power-law index $p$. The results of these fits are given in Table~\ref{table.ulx}, along with fits for the same model to the combined 2017 and the combined 2011 and 2017 spectra. The 2011$+$2017 fit is not formally acceptable (with a $p$-value of 0.02), but the parameters are basically consistent with the results from the individual epochs: $N_{\text{H}} = 1.9\pm0.1 \times 10^{21}$~cm$^{-2}$, $T_{\text{in}} = 1.67\pm0.06$~keV, and $p=0.538\pm0.007$. However, since so many of the counts come from the 2011 data, a fairer comparison is between the combined 2017 spectra and the 2011 spectra. This fit suggests that the disk temperature has increased slightly since 2011, but most of the individual epochs during 2017 are consistent with the 2011 parameters. 

The updated light curve is shown in Figure~\ref{figure.ulx_fluxes}. The ULX is clearly variable on a timescale of about a day, but during the 2017 observations it remained close to the $4-5\times 10^{39}$~erg~s$^{-1}$ level from 2012. Thus, the ULX has settled to a new level instead of returning to quiescence. Meanwhile, the spectrum has remained about the same.

\begin{deluxetable*}{lcccccc}
\tablenum{5}
\tabletypesize{\scriptsize}
\tablecaption{ULX {\sc TBabs(diskpbb)} Fits}
\tablewidth{0pt}
\tablehead{
\colhead{Epoch} & \colhead{$N_{\text{H}}$} & \colhead{$T_{\text{in}}$} & \colhead{$p$} & \colhead{$F_X$} & \colhead{$\chi^2$} & \colhead{dof} \\
\colhead{} & \colhead{($10^{22}$~cm$^{-2}$)} & \colhead{(keV)} & \colhead{} & \colhead{($10^{-13}$~erg~s$^{-1}$~cm$^{-2}$)} & \colhead{} & \colhead{}
}
\startdata
2011-08-25 & $0.20\pm0.01$ & $1.66\pm0.07$ & $0.54\pm0.02$ & $11\pm2$ & 1665.5 & 1626 \\
2017-01-27 & $0.16\pm0.04$ & $1.7\pm0.2$ & $0.59\pm0.06$ & $4\pm1$ & 359.6 & 343 \\
2017-01-29 & $0.18\pm0.03$ & $1.7\pm0.2$ & $0.57\pm0.04$ & $4\pm2$ & 231.5 & 258 \\
2017-02-19 & $0.17\pm0.05$ & $1.7\pm0.3$ & $0.57\pm0.04$ & $3\pm1$ & 206.8 & 219 \\
2017-02-23 & $0.18\pm0.02$ & $2.2\pm0.3$ & $0.52_{-0.01}^{+0.05}$ & $3\pm1$ & 212.0 & 235 \\
2017-02-25 & $0.14\pm0.05$ & $1.5\pm0.2$ & $0.60_{-0.06}^{+0.03}$ & $3\pm1$ & 355.1 & 370\\
2017 All      & $0.18\pm0.02$ & $1.8\pm0.1$ & $0.56\pm0.02$ & - & 1462.9 & 1445 \\
2011$+$2017  & $0.19\pm0.01$ & $1.67\pm0.06$ & $0.538\pm0.008$ & - & 3225.6 & 3066 
\enddata
\tablecomments{\label{table.ulx} Best-fit parameters for the {\sc TBabs(diskpbb)} model in different epochs. The error bars are the 90\% confidence intervals from MCMC sampling. The final two rows list the results for the combined 2017 and the combined 2011-2017 spectra. 
}
\end{deluxetable*}

\subsection{Model and Systematic Effects}

We investigate the impact of systematic effects, abundance tables, and model choices on our measurements of temperature and metallicity. Table~\ref{table.bias} summarizes the impact of these effects and choices on the metallicity, as quantified through a $\Delta Z/Z_{\odot}$ to be added to the fiducial value.

\subsubsection{Temperature}
The temperature of the dominant component, $kT_1$, is insensitive to the low resolution, choice of model, and most systematic effects because the line complexes whose ratios determine the temperature are well defined (Figure~\ref{figure.halo_spectra}). Indeed, even when omitting the strong oxygen lines, we find that $kT_1=0.22\pm0.02$\,keV, which is consistent with our fiducial $kT_1 = 0.20\pm0.01$\,keV. When using the {\sc mekal} model \citep{mewe85,mewe86,liedahl95} instead of {\sc apec}, we find $kT_1 = 0.21\pm0.01$\,keV. There is no statistical preference for a 3T model, which is at most a small perturbation to the 2T model. The temperature could be biased if the Galactic foreground emission lines are incorrectly measured, but the background spectra across all epochs are fit well by a single model, so the bias is smaller than the statistical error. Likewise, temperatures measured with the MOS, pn, and ACIS detectors are all consistent within the statistical error bars.

In contrast, the hotter component is weaker and more susceptible to low resolution or systematic effects. For $0.3 < kT < 1.0$~keV, the main emission lines are a cluster of Fe-L lines between $E=0.7-1.2$~keV, which are unresolved by the EPIC or ACIS detectors and appear as a bump. The wavelength of the bump peak determines the temperature. Secondary factors include the relative strength of the \ion{O}{8} line and other lines between $E=1-2$\,keV. We measure $kT_2 = 0.77_{-0.06}^{+0.07}$~keV in the outer halo and $kT_2 = 0.68\pm0.04$~keV in the inner halo (90\% error bars). The difference exceeds the statistical error bars, but after investigating various spectra and model components, we found that it can be entirely explained by the pn spectrum in the inner halo, which demands a lower $kT_2 = 0.67\pm0.04$~keV. The MOS $kT_2 = 0.74\pm0.08$ in the same region. This difference persists when using only the 2011 or only the 2017 spectra, so the ULX is not responsible. 

At the energy resolution of the pn, spectra with temperatures $0.4 < kT < 0.7$~keV monotonically increase towards a peak near $E=0.8-0.9$~keV. When combined with a cooler spectrum, it blends with the strong \ion{Fe}{17} line at 0.82~keV. This blend is less severe in the MOS, which can lead different best-fit temperatures. We did not find any issues with the response matrices or obvious differences in the extraction regions, so it is not clear which is correct, but the inner-halo MOS parameters agree with the outer-halo MOS and pn parameters, so we cannot conclude that the temperature differs. 

\subsubsection{Metallicity}
\begin{figure}
\begin{center}
\includegraphics[width=1\linewidth]{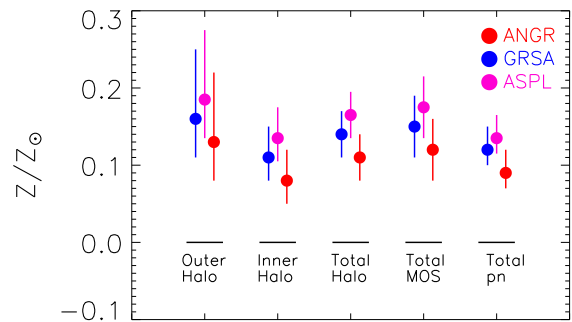}
\vspace{-1.25cm}
\caption{Metallicities from the 2T, PL free model for the outer halo, inner halo, and whole halo, respectively. The best-fit values for the MOS and pn alone are also shown. The red, blue, and magenta points correspond to the \citet{anders89} (default in \textit{Xspec}), \citet{grevesse98}, and \citet{asplund09} Solar abundance tables, respectively.}
\label{figure.abundcompare}
\end{center}
\end{figure}

At $kT = 0.2$\,keV, the bremsstrahlung continuum above about 0.8\,keV is weak, so the metallicity measurement depends sensitively on the oxygen lines and the pseudo-continuum at $E<0.5$\,keV, and between about $0.7<E<0.8$\,keV. The low-energy continuum is most susceptible to systematic issues, such as residual soft-proton flares, spatial variation in the quantum efficiency of the MOS cameras below $E<0.4$\,keV, build-up of contaminants on the MOS, change in energy resolution in the pn over time, and small uncertainties in the effective area for both cameras. Figure~\ref{figure.abundcompare} shows the differences between metallicites measured in the outer and inner halo regions, between the MOS and pn spectra only, and for several Solar abundance tables.

\paragraph{Residual Flares} Flares that are too weak to detect in the background light curve contribute some counts. The spectrum is very soft and unabsorbed, so the thermal continuum below $E=0.5$~keV may be overestimated, which would bias $Z/Z_{\odot}$ towards lower values. This effect is most pronounced in the pn. They are clearly present in the data, since the 2017 pn spectra have a higher flux below $E=0.5$~keV than the 2011 spectrum, whereas the \ion{O}{7} line strength is about the same. These flares are accounted for in the spectral modeling of the background, but if the proton flux is underestimated then the metallicity would be biased. For example, when fitting the pn spectra from the outer halo with the 2T model but without accounting for residual flares, we find $Z/Z_{\odot} = 0.06$. 

We estimated the bias from residual flares by varying the fitting bandpass and using the average \XMM{} flaring properties. First, we adopted a pn fit bandpass of 0.45-5~keV, while keeping the MOS at 0.3-5~keV, to reduce the impact of flares. This approach relies on the MOS continuum to determine the metallicity. We find a best-fit $Z/Z_{\odot} = 0.20_{-0.07}^{+0.11}$ in the outer halo and $Z/Z_{\odot} = 0.16\pm0.04$ in the whole halo, for $\Delta Z/Z_{\odot} \approx +0.03$. However, this approach discards the signal needed to measure the metallicity in the first place. Alternatively, we explicitly fitted the residual soft proton flares in the source spectrum (as opposed to scaling from the background spectrum) using a pn fit bandpass of 0.3-5~keV. We used the same broken power-law model from ESAS and find $Z/Z_{\odot} = 0.22_{-0.12}^{+0.65}$, or $\Delta Z/Z_{\odot} = +0.06$, but with very poor constraints. This approach leads to a maximal contribution from the soft proton flares, and so this is an upper bound.

We then estimated the total count rate expected from residual flares using the average number of flares as a function of intensity \citep{carter08}, the proton vignetting function, our fitting region, the detection threshold for the lightcurves we used based on the ESAS approach, and the duration of our exposures. We adopted a power-law index of $\Gamma=2.0$ for the flare energy spectrum. The model flux from this approach was somewhat lower than that inferred from the background region fits for most of the 2017 epochs, but higher than for 2011, and it did not have a significant effect on the metallicity: $Z/Z_{\odot} = 0.18_{-0.06}^{+0.09}$, or $\Delta Z/Z_{\odot} = +0.02$. 

The bias from residual proton flares is thus at most $\Delta Z/Z_{\odot} = +0.06$, and is likely closer to +0.02. This could be measured more accurately by using \Chandra{} data, but the existing data sets have too few counts below $E<0.5$~keV do to this. 

\begin{figure}
\begin{center}
\includegraphics[width=0.95\linewidth]{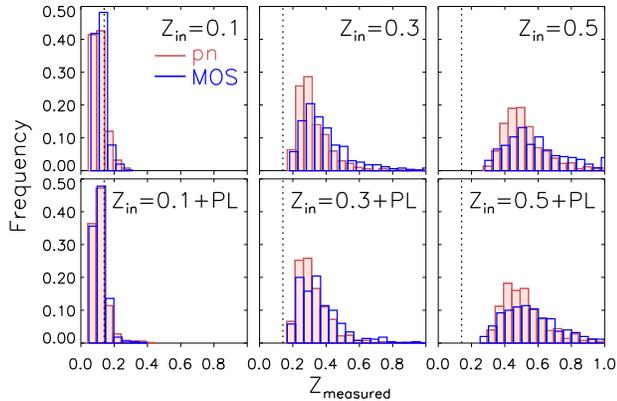}
\caption{The distribution of measured $Z/Z_{\odot}$ values from simulated spectra with different input $Z/Z_{\odot}$ and matched $S/N$ to the \XMM{} data. The top row shows simulations from a thermal model only, while the bottom includes a power-law component. The red, filled histograms and blue, outline histograms correspond to the MOS and pn, respectively. The vertical dashed line indicates our fiducial $Z/Z_{\odot} = 0.14$. These plots show that the metallicity is more likely to be overestimated than underestimated by shot noise, and that $Z/Z_{\odot} \ge 0.3$ is strongly disfavored if the measured metallicity is as low as 0.14.}
\label{figure.zmeas}
\end{center}
\end{figure}

\paragraph{Low energy calibration} 
A detailed investigation of the $E<0.5$\,keV calibration is beyond the scope of this work, but 
based on the EPIC calibration documents\footnote{https://www.cosmos.esa.int/web/xmm-newton/calibration-documentation} we simulated MOS and pn spectra to determine the impact of uncertainties in the effective area calibration using the {\sc corrarea} option when generating the ancillary response. We used the fiducial model as an input, but varied $Z/Z_{\odot}$ from 0.1 to 0.5. 
Based on fits to these spectra, we estimate that the calibration-related bias is $\Delta Z/Z_{\odot} < +0.06$, with a typical value of $\Delta Z/Z_{\odot} = +0.02$. The lower value is obtained when using an \textit{ad hoc} correction of fitting spectra explicitly including a multiplicative factor to account for the cross-calibration uncertainty\footnote{See  https://heasarc.gsfc.nasa.gov/docs/xmm/esas/cookbook/xmm-esas.html}. Figure~\ref{figure.abundcompare} shows that the values obtained from the MOS and pn separately do not differ much.

\paragraph{Spectral Resolution}
The metallicity can be incorrectly measured due to low spectral resolution. We used the MOS and pn responses for our spectra to simulate spectra with the same $S/N$, using an absorbed isothermal plasma with $N_{\text{H}} = 6.8\times 10^{20}$\,cm$^{-2}$, $kT=0.20$\,keV, and $Z/Z_{\odot} =$0.1, 0.3, and 0.5 as input models. We simulated 1000 spectra for each $Z/Z_{\odot}$, then fit the spectra with the same model (with $N_{\text{H}}$ fixed). The histograms of the measured $Z/Z_{\odot}$ for the MOS and pn are shown in Figure~\ref{figure.zmeas}. We then repeated the experiment including a power-law component with $\Gamma=1.4$ and the same flux as in the best-fit 2T model. These histograms are also shown in Figure~\ref{figure.zmeas}, where a vertical dashed line indicates the best-fit $Z/Z_{\odot}$ for the whole halo. The simulations indicate that the power law does not substantially affect the ability to measure the metallicity, and show that random fluctuations typically lead to an overestimate of the metallicity, as opposed to the other effects that lead to an underestimate; our measured value of $Z/Z_{\odot} = 0.14$ is very unlikely if the true $Z/Z_{\odot} \ge 0.3$. Near the measured value, we estimate $\Delta Z/Z_{\odot} = -0.02$ based on these simulations.

We also verified that spectral resolution and pile-up do not significantly affect the results by extracting spectra using only single (pattern=0) events for the outer halo, which represent more than 75\% of events detected at $E<0.7$\,keV. We repeated the 2T model fits and found good agreement with our reported results  (Table~\ref{table.spectra}).

\begin{deluxetable}{lcc}
\tablenum{6}
\tabletypesize{\scriptsize}
\tablecaption{Metallicity Bias}
\tablewidth{0pt}
\tablehead{
\colhead{Source} & \multicolumn{2}{c}{$\Delta Z/Z_{\odot}$} \\ 
}
\startdata
\multicolumn{3}{c}{Systematic Error}\\
\hline
Residual Flares & $+0.02$ & ($<$ $+0.06$) \\
$A_{\text{eff}}$ Cross-calibration & $+0.02$ & ($<$ $+0.05$) \\
Spectral resolution & $-0.02$ & \\
Resonant scattering       & $+0.02$ & ($<$ $+0.05$) \\
\hline
\multicolumn{3}{c}{Abundance Table}\\
\hline
\citet{grevesse98}\tablenotemark{f} & $+0.00$ & \\
\citet{anders89}  & $-0.04$ & \\
\citet{asplund09} & $+0.05$ & \\
\citet{wilms00} & $+0.03$ & \\
\hline
\multicolumn{3}{c}{{\sc Xspec} Model}\\
\hline
{\sc phabs}\tablenotemark{f} & $+0.00$ & \\
$N_{\text{H}}$ value & +0.0 & ($<$ $-0.06$) \\
{\sc TBabs} & $+0.02$ & \\
2T {\sc apec}\tablenotemark{f} & $+0.00$ & \\
{\sc mekal} & $-0.05$ & \\
1T {\sc apec} & $-0.01$ & 
\enddata
\tablenotetext{f}{Fiducial Model}
\tablecomments{\label{table.bias} The adjustments to the metallicity value resulting from possible systematic effects and model choices are added to the value for the fiducial model, $Z/Z_{\odot} = 0.14\pm0.03$. The issues are discussed in the text.
}
\end{deluxetable}

\paragraph{Galactic column}
The effect of absorption is strongest below $E<0.5$~keV, and the range in reported Galactic $N_{\text{H}}$ values \citep{kalberla05,dickey90} leads to a $-0.06 < \Delta Z/Z_{\odot} < 0$ for $7.8\times 10^{20} > N_{\text{H}} > 6.5\times 10^{20}$~cm$^{-2}$, relative to the fiducial value of $N_{\text{H}} = 6.8\times 10^{20}$\,cm$^{-2}$ for the {\sc phabs} model. 

\paragraph{{\sc apec} vs. {\sc mekal}}
We compared the fiducial {\sc apec} results to those from the {\sc mekal} model. We used the {\sc mekal} option to calculate the model instead of interpolating from existing tables, and find $\Delta Z/Z_{\odot} = -0.05$ for the fiducial \citet{grevesse98} values. 

\paragraph{Solar Abundance Ratios}
We compared results when using different Solar abundance tables from \citet{anders89}, \citet{asplund09}, and \citet{wilms00}, relative to the fiducial \citet{grevesse98} values.  This led to $\Delta Z/Z_{\odot} = -0.04$, +0.05, and +0.03, respectively (Figure~\ref{figure.abundcompare}). The {\sc TBabs} absorption model requires the \citet{wilms00} absorption table, and when we use this, $\Delta Z/Z_{\odot} = +0.02$. 

There are several important lines between $0.3 < E < 0.5$\,keV at $kT=0.2$\,keV, most notably the \ion{C}{6} line at 0.37\,keV and the \ion{N}{6} line at 0.42\,keV, so \citet{kuntz10} examined the possibility that incorrect carbon and nitrogen abundances could explain the low metallicity found in parts of M101. We repeated their procedure for the NGC~891 spectra, using a {\sc vapec} model, but allowing the carbon and nitrogen abundances to float and fixing the metallicity at solar. A good fit requires that these abundances be 3-5 times higher than the fiducial values, which is consistent with \citet{kuntz10} and inconsistent with supernova yields and the other abundances. As far as we are aware, no serious problem with the Solar carbon and nitrogen abundances has been found. 

\paragraph{Optical Depth}
Finally, we examined the effect of resonant scattering in the oxygen lines, which would reduce the line strength relative to the continuum near the disk and lead to an underestimate of the metallicity. We have assumed that the halo is optically thin (the coronal approximation), but \citet{li_yunyang17} showed that the Milky Way's hot halo is optically thick in the 0.573\,keV (21.6\AA) resonance line of \ion{O}{7} and the 0.654\,keV (18.9\AA) line of \ion{O}{8}, with $\tau \gtrsim 1$ towards the Galactic center. Since the hot halo of NGC~891 is somewhat brighter than that of the Milky Way, one expects $\tau > 1$ in the inner regions, and non-negligible $\tau$ even in the outer halo, since $\tau$ declines as $1/r$.

We estimated $\tau$ for the cases of solar metallicity and the best-fit metallicity in both the inner and outer halo. We assume a cylindrical ring geometry for the halo, with $R_2 = 7$~kpc, based on the location of the bright X-ray emission, and a variable $R_1$. The volume is then $V = \pi (R_2^2-R_1^2)\times 2h$, where $h=2.2$~kpc for the inner halo and $6.5$~kpc for the outer halo. The factor of 2 is because the {\sc apec} model normalization is for both sides of the midplane. We obtain the mean density from the {\sc apec} normalization quantity, which is
\begin{equation}
{\rm norm} = \frac{10^{-14}}{4\pi d_L^2}\int n_e n_H dV.
\end{equation}
Then, the optical depth for \ion{O}{7} is 
\begin{equation}
\tau/L = \kappa(s) = \bar{n} A_{\text{O}} X_{\text{ion}} \sigma \approx \bar{n} A_{\text{O}} X_{\text{ion}} \times 0.015 f b^{-1} \lambda,
\end{equation}
where $L$ is the line-of-sight distance in cm, $A_{\text{O}}$ is the oxygen abundance, $X_{\text{ion}}$ is the ion fraction, $\sigma$ is the absorption cross-section, $f = 0.696$ is the oscillator strength, $b$ is the Doppler $b$ parameter, and $\lambda = 21.602\times 10^{-8}$~cm is the transition wavelength. $b=\sqrt{b_{\text{turb}}^2 + b_{\text{thermal}}^2}$, where $b_{\text{thermal}} \approx 1.29\times 10^4 \sqrt{T/m_{\text{O}}}$~cm~s$^{-1}$. For $kT=0.20$~keV, $X_{\text{ion}} = 0.5$ and $b_{\text{thermal}} \approx 50$~km~s$^{-1}$. As mentioned above, there are different values for the Solar oxygen abundance in the literature. For consistency with the \citet{grevesse98} abundance table, we use $\log \text{O/H} + 12 = 8.83$. Finally, we assume $b_{\text{turb}} = 75$~km~s$^{-1}$, based on hydrodynamic models of galaxy halos \citep{crain10,nuza14}.

The {\sc apec} normalizations for the fiducial model are $\text{norm}_{\text{inner}}=3.5\times 10^{-4}$ and $\text{norm}_{\text{outer}} = 3.4\times 10^{-4}$ for the inner and outer halo, respectively. This model has $Z/Z_{\odot} = 0.14$. We also fitted the spectra for the inner halo only with $Z/Z_{\odot} \equiv 1.0$, which leads to $\text{norm}_{\text{inner}} = 3.4\times 10^{-5}$; we note this is a significantly worse fit. In the fiducial model, the optical depth for $R_1 = 0$~kpc is $\tau_{\text{inner}} \approx 1.8$ and $\tau_{\text{outer}} \approx 1.0$. The low metallicity $kT=0.20$~keV component must have a large volume-filling factor, so these are reasonable approximations. Moreover, they are independent of the size of $R_2$. For $Z/Z_{\odot} \approx 1$, $\tau_{\text{inner}} \approx 5$. In this case, it is more likely that the hot gas is outflowing material that forms a ring of hot gas above the disk, based on inferences for where star formation occurs \citep{boettcher16}. If we set $R_1 = 4$~kpc, then $\tau_{\text{inner}} \approx 2.5$. 

These estimates should be interpreted cautiously, since we do not know the actual geometry of the gas, nor its turbulent velocity (if the gas is quiescent, the optical depth is higher by a factor of $\sim$2). Nonetheless, they suggest that optical depth will be a mild effect, especially in the outer halo. A significant amount of the measured \ion{O}{7} flux is contributed by the forbidden and intercombination lines, which are optically thin. In the inner halo, absorption due to the \ion{H}{1} halo is likely more important. We tested the impact by fitting the spectra while ignoring the \ion{O}{7} complex. Since $\tau$ is lower for \ion{O}{8} than \ion{O}{7}, this should have a higher metallicity if the coronal approximation biases our measurement. The inner-halo $Z/Z_{\odot} = 0.16_{-0.05}^{+0.08}$ when disregarding the $E=0.5-0.6$~keV band, which is slightly higher than the original $Z/Z_{\odot} = 0.11_{-0.03}^{+0.04}$ and in better agreement with the outer halo $Z/Z_{\odot} = 0.16_{-0.05}^{+0.09}$, but also consistent with the original within the 90\% error bars. This suggests that the optical depth in the \ion{O}{7} line biases the metallicity by at most $\Delta Z/Z_{\odot} < +0.05$ for the \textit{whole} halo.

These effects cannot all contribute to $\Delta Z/Z_{\odot}$ maximally, and their relative magnitude will differ between regions. For instance, for residual proton flares to bias the measurement by $\Delta Z/Z_{\odot} = 0.06$, the reduction in the \ion{O}{7} line strength by resonant scattering could not lead to $\Delta Z/Z_{\odot} +0.05$ and be consistent with the data. For the fiducial fit to the whole halo, we estimate that $Z/Z_{\odot} = 0.14 \pm 0.03$~(stat)~$^{+0.08}_{-0.02}$~(sys), not including model choices ({\sc apec} vs. {\sc mekal}, or the abundance table). The systematic uncertainty exceeds the statistical uncertainty, so NGC~891 has reached the limiting $S/N$ for measurements with \XMM{}.

\section{Discussion}
\label{section.discussion}

We now consider the origin of the hot gas, the origin of the nonthermal emission, how the hot gas fits into the multi-phase halo around NGC~891, and the implications of our results for measurements made at lower $S/N$ around other spiral galaxies. 

\subsection{Origin of the Hot Gas}

The visible hot gas could be outflowing gas, the inner part of an extended hot halo of gas accreted from the intergalactic medium, or a combination of the two. We consider the case for each scenario, based on the measurements and derived quantities in Table~\ref{table.halo_properties}. 

\subsubsection{Temperature}
Extraplanar gas around most spiral galaxies can be fit well by a 2T model, and NGC~891 is no exception. However, the comparatively higher $S/N$ in these \XMM{} data allow us to draw two additional conclusions: the 2T model is ``real'' in the sense that there are two distinct components with different temperature distributions (a 3T or 4T model collapses to the 2T case and charge exchange is ruled out as a significant contributor), and these components have a different spatial distribution. Figure~\ref{figure.tempmap} indicates that the hotter gas is primarily located directly above the star-forming regions where there are filamentary structures in the \Chandra{} map (Figure~\ref{figure.chandra_filaments}). The relative temperature map is based on an isothermal model, but our fits to the inner and outer halo, and to the disk and filament regions corroborate the basic picture: most of the hotter, $kT_2 = 0.6-0.8$~keV gas is located closer to the disk.

The temperature does not clearly favor one origin scenario. The temperature map strongly suggests that there are ongoing outflows, and outflowing gas is likely to have a temperature exceeding the virial temperature, $kT_{\text{vir}} \approx 0.18$~keV for $v_c = 212$~km~s$^{-1}$. However, the dominant component at $kT_1 = 0.20$~keV could be cooled, relaxed ejecta or virialized infall, both of which would lead to a similar morphology and temperature. 

\subsubsection{Metallicity}
The fiducial metallicity of the dominant $kT_1 = 0.2$~keV component is $Z/Z_{\odot} = 0.14\pm0.03$, with $Z/Z_{\odot} < 0.2$ likely (Section~\ref{section.spectra}). This is consistent with an accreted hot halo, based both on models \citep[e.g.,][]{shen10,crain10} and measurements \citep{aguirre08} of the intergalactic and halo metallicity at $z<1$. In this case, the visible X-rays come from the inner part of a more extended, massive hot halo. Such a halo would trap metals from SNe-driven outflows close to the disk \citep{tenorio-tagle02}, so enrichment of the extended halo would be slow. 

In contrast, we expect the halo gas to have a metallicity $Z/Z_{\odot} > 0.5$ in the outflow scenario. For a late-type galaxy such as NGC~891, the Type~II SN (SN~II) rate exceeds the Type~I (SN~I) rate. The SN~II rate can be estimated from the SFR$=3.8 M_{\odot}$~yr$^{-1}$, which implies a rate $R_{\text{SN II}} \approx 0.03$~yr$^{-1}$ if we assume steady star formation, the \citet{kroupa02} initial-mass function, and that SN~II progenitors are at least $M>8 M_{\odot}$. This agrees with the estimate from the stellar mass \citep[$M_* \approx 5\times 10^{10} M_{\odot}$;][]{li13a} and spectral type (Sb) from Table~2 in \citet{mannucci05}, $R_{\text{SN II}} \approx 0.04$~yr$^{-1}$. In contrast, the SN~I rate using the same approach is $R_{\text{SN I}} \approx 0.01$~yr$^{-1}$. Thus, most of the SNe feedback will come from SN~II. 

In this case, most of the outflowing gas will come from superbubbles \citep{weaver77,maclow89,silich05,roy13,baumgartner13}, since the contributions of multiple SNe are needed to break out of the thin disk. The extraplanar dust chimneys in NGC~891 indicate that this process is indeed at work in NGC~891 \citep{howk00}. Bubbles filled with hot gas form around star clusters as a result of SNe and stellar wind heating, then expand into the ISM and sweep up a neutral shell. The shell decelerates as it expands, but once it breaks out of the thin disk it accelerates down the pressure gradient and breaks up, leaving a chimney. The hot gas then flows out of as a pressure-driven wind.

The hot gas in superbubbles is well mixed \citep{tenorio-tagle02}. When the bubble has just formed, the hot gas (which comes from stellar winds and SNe) has a super-solar metallicity. As the bubble grows, this material evaporates neutral gas from the bubble wall. The hot temperature and corresponding large sound speed promote rapid mixing, which dilutes the hot gas and reduces its metallicity to some value intermediate between the SNe ejecta and the surrounding ISM \citep[see][]{silich01}. In principle, this can lead to low metallicity by the time of breakout, but most of the star formation in NGC~891 (and the X-ray emission around it) is located within a galactocentric radius of $R<4$~kpc. In this region, other, face-on spiral galaxies of a similar mass have near-Solar gas-phase and photospheric metallicities \citep{kudritzki15}. Thus, we would expect a pre-breakout superbubble to have around $0.5 < Z/Z_{\odot} < 2$. An extremely deep X-ray map of the face-on spiral M101 bears out this expectation, as \citet{kuntz10} report an average metallicity for the hot ISM of $Z_{\text{Fe}}/Z_{\odot,\text{Fe}} \approx 0.5$ \citep[they fix the oxygen abundance at the gas-phase $Z_{\text{O}}/Z_{\odot,\text{O}} = 0.75$ from][]{bresolin07}. Individual superbubbles with sufficient energy injection to break out will have a higher metallicity than this average. 

The metallicity of the halo gas could be low if the metals are segregated in a hotter or cooler phase. For example, most of the energy and metals in galactic superwinds are thought to be in a very hot ($T>10^7$~K, $kT>1$~keV) phase that has low density, and thus low emissivity, due to its rapid speed \citep[$v>1000$~km~s$^{-1}$;][]{veilleux05,heckman17}. However, NGC~891 does not have a superwind, and the winds from individual superbubble breakout regions will be cooler and slower \citep{maclow89,roy13,baumgartner13}, with velocities less than a few hundred km~s$^{-1}$ and temperatures in the soft X-ray regime. Alternatively, metals could clump and condense out of the hot gas before breakout. Similar condensation from poorly mixed gas has been proposed to explain the low metallicity seen in some early-type galaxies or spiral bulges \citep{tang09,pipino11}. However, superbubbles are qualitatively different from the SN~I winds proposed in those systems, and in particular the consecutive SNe in a small bubble leads to efficient mixing and the destruction of any Fe-rich dust that may form. Breakout into the halo occurs at latest when most of the SNe have occurred \citep{baumgartner13}, and condensation of metal-rich regions would need to occur after this period when there is no further shock heating \citep{tenorio-tagle02}.

There is a family of models in which cosmic rays from SNe help to drive galactic outflows \citep{breitschwerdt91}. Cosmic rays contain a large amount of momentum and can couple to both cool and hot gas. However, this cannot easily reduce the metal content in the hot gas, nor does it heat or cool it, except insofar as the hot gas may propagate farther outwards. In contrast to pressure-driven superwinds, which are accelerated to $v>1000$~km~s$^{-1}$ near the starburst, cosmic-ray driven winds accelerate down the pressure gradient out of the disk and have velocities of at most a few hundred km~s$^{-1}$ near the disk (for a Milky Way-like SFR). These models do not obviously change the expectation of near-Solar metallicity for hot, outflowing gas. 

\subsubsection{Spatial Distribution}
As in many other spiral galaxies \citep{strickland04,tullmann06,li13a}, the emission from the $kT_1 = 0.2$~keV gas is brightest above star-forming regions and forms a thick disk. There is filamentary emission in the \Chandra{} image above these regions (Figure~\ref{figure.chandra_filaments}) and a large filament extending to $\sim$15~kpc above the galactic center on the west side.

This basic structure is expected in the outflow scenario, as superbubble winds are powered by massive star formation. Newly formed outflows from active chimney regions would appear filamentary, whereas expelled gas will either relax and virialize or cool and migrate to larger galactocentric radii \citep{melioli09}, depending on its temperature. In either case, the collective outflows from above the star-forming region of the disk would lead to an exponential disk atmosphere \citep{wang10}. 

By itself, an accreted hot halo is unlikely to have this morphology. Instead, we expect a spheroidal power-law radial profile, which has been observed at larger scales \citep{miller15,li17}. However, the ambient hot gas will be affected by outflows and the presence of cold gas, so we briefly explore whether a disk-like morphology is consistent with the inner part of an accreted halo around a star-forming disk. 

Outflows will affect the ambient halo in at least two ways. First, winds from individual superbubbles will ram into the hot halo, shock, and decelerate. This interaction will form bubbles of hotter wind gas surrounded by shocked or compressed halo gas. This both traps outflowing metals and brightens the ambient hot halo in the vicinity of the star formation. Second, outflowing gas drags magnetic fields into the halo \citep[as evidenced by the ``X-shaped'' fields in radio halos;][]{krause09}. The average magnetic pressure around star-forming galaxies can exceed the average thermal pressure, as in the edge-on, star-forming galaxy NGC~4631 \citep{irwin12b}. This would also tend to compress ambient hot gas in the vicinity of star formation. The magnetic scale heights of radio halos tend to be large ($>4$~kpc) and the synchrotron emission can be described by an exponential disk, so we would expect compression to imprint a disk-like morphology onto the ambient hot halo. Since both the outflow strength and the amount of cosmic rays leaving the disk depends on the SFR, the brightening of the ambient hot halo would be related to the SFR. 

In the compression scenario, we would expect to see shells of hot (and maybe cold) gas. Such shells are seen in NGC~5775 \citep{li08} and NGC~4631 \citep{irwin12b}, and in NGC~891 there is a hint of such a shell in the soft \XMM{} image at the top of the large filament (Figure~\ref{figure.xmm_images}). However, there is no compelling evidence for shells in the \Chandra{} image, although it is difficult to distinguish them from filaments and it is hard to detect individual shells in projection. 

NGC~891 has an X-shaped magnetic field and a large radio halo \citep{allen78,hummel91,wiegert15}. The mean $B = 7$~$\mu$G at $1 < z < 3$~kpc above the disk \citep{mulcahy18} implies $P_B \approx 2\times 10^{-12}$~dyne~cm$^{-2}$. Meanwhile, the mean $P_{\text{th}} \approx 7\times 10^{-13}$~dyne~cm$^{-2}$, so the magnetic pressure is about three times higher. Magnetic compression of an ambient hot halo therefore appears energetically plausible, but magnetohydrodynamic simulations of superbubble breakouts in the presence of a hot halo are needed to explore this scenario.

\begin{figure}
\begin{center}
\includegraphics[width=1\linewidth]{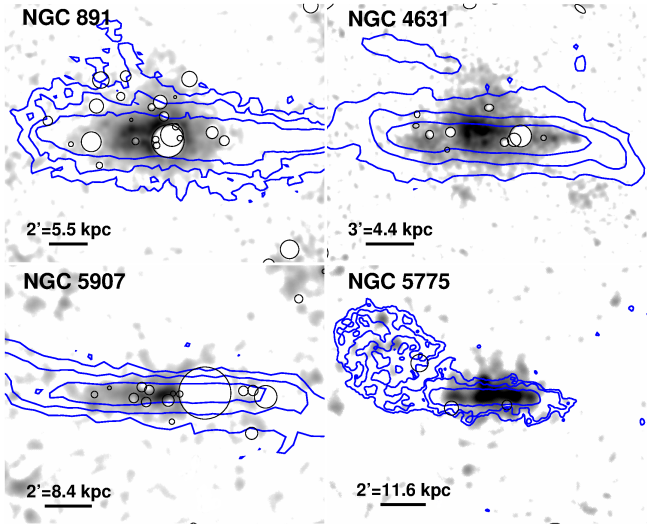}
\caption{\ion{H}{1} contours (blue) on 0.4-1.4~keV (greyscale) \XMM{} images of NGC~891, NGC~4631, NGC~5907, and NGC~5775. The contours are at 3, 10, and 30$\times 10^{20}$~cm$^{-2}$. Point source masks are shown as black circles. Almost all of the extraplanar X-ray emission, which has a smaller scale height than expected from the temperature, comes from within the thick \ion{H}{1} halo, but it does not seem to follow the contours, which we would expect if most of the X-ray luminosity comes from cooling in the wake of cool clouds \citep{marinacci10}.}
\label{figure.n891_HIcompare}
\end{center}
\end{figure}

The other way that cold gas can shape the hot emission is by stimulating cooling \citep{armillotta16,fraternali17}. This occurs in the wakes of cool clouds traveling through the hot halo because the outer layers of the clouds evaporate and mix. The cooling time declines both because some thermal energy goes into heating the cool gas and because the mixture has a higher density, which increases the luminosity. A thick \ion{H}{1} disk would then also lead to a thick X-ray disk. However, Figure~\ref{figure.n891_HIcompare} shows that the \ion{H}{1} morphology is not closely connected to the X-ray emission for NGC~891 and three other edge-on galaxies with different SFR and sizes. Stimulated cooling thus appears to be a secondary effect.

\subsubsection{Summary and Proposed Scenario}
The existence of super-virial ($kT_2 = 0.6-0.8$~keV) gas that is concentrated near the disk and the shape of the X-ray emission, including filamentary structures, shows that some of the hot gas originated in outflows. However, the low metallicity of the dominant, virialized component implies that most of the gas did not. We therefore propose a scenario in which stellar feedback flows into the inner region of an ambient, extended hot halo, becomes trapped, cools, and falls back without substantially enriching its surroundings. The compression of the ambient gas by the galactic fountain outflows leads to a halo that is substantially brighter than expected from the extended hot halo alone, and which is brightest around the star-forming regions. 

Additional observations could falsify this model, although it is unlikely that additional \XMM{} or \Chandra{} data could do so due to the poor spectral resolution. Specifically, the metallicity of the hotter component must be near-Solar in this model; if not, then outflowing hot gas may indeed find a way to hide its metal content. The \textit{Resolve} calorimeter on the planned \textit{X-ray Imaging and Spectroscopy Mission} \citep{kelley16} will resolve the individual line complexes and lead to more robust temperature and metallicity measurements in both components. Alternatively, the proposed \textit{Advanced X-ray Imaging Satellite} \citep{mushotzky18} or \textit{Lynx} High Definition X-ray Imager \citep{gaskin18} would be able to detect the extended hot gas around numerous $L_*$ spiral galaxies to $r<100$~kpc, which would directly confirm the existence of extended, virialized hot halos.

\subsection{The Galactic Fountain and Cool Gas}

The hot gas co-exists with a large amount of cool gas: there is a remarkably massive \ion{H}{1} halo with $M_{\text{cold}} = 1.2\times 10^9 M_{\odot}$ that extends to $z>10$~kpc \citep{oosterloo07} that contains at least $6\times 10^6 M_{\odot}$ of dust \citep{bocchio16,hk16}, a nearly-as-massive $< 10^9 M_{\odot}$ halo of extraplanar molecular gas up to 1-2~kpc above the disk \citep{garcia-burillo92}, and warm gas associated with diffuse ionized gas \citep{rand94}. The cold gas mass is at least an order-of-magnitude greater than the hot gas mass, and thus cannot condense from the hot halo in the steady-state \citep{hk13}. If only the $kT_2 \sim 0.7$~keV gas is from outflows, the difference is at least a factor of 100. 

The origin of the cold gas is not clear, but the preponderance of evidence favors ejection from the disk. The metallicity of this material is $Z/Z_{\odot} > 0.4$ (Qu, Bregman \& Hodges-Kluck, 2018), the dynamics favor a scenario in which most of the material was ejected but with some braking by an accreted halo \citep{oosterloo07}, and the dust-to-gas ratio is consistent with a typical $L_*$ spiral galaxy disk. Meanwhile, most of the gas is unlikely to come from interaction with a satellite. Although there is an \ion{H}{1} filament pointing in the direction of a  companion galaxy, its mass is too small to account for much of the halo gas mass. A deep examination of the stellar halo found no evidence that the stellar streams around NGC~891 are associated with this companion galaxy or any recently disrupted satellites \citep{mouhcine10}. 

This would seem to favor a stellar feedback origin for the hot gas (in contrast to what we argued above), but the amount of cold mass is difficult to reconcile with any model. There is clearly enough kinetic energy from SNe to account for a few$\times 10^9 M_{\odot}$ in the halo \citep{fraternali06}, but the mass-loading rate is uncharacteristically high. Assuming that the halo passage time for a cold cloud ejected from the disk is about the orbital time in the star-forming region ($\sim$200~Myr), about $10 M_{\odot}$~yr$^{-1}$ of cool gas must be ejected from the disk to sustain the \ion{H}{1} and H$_2$ halos in the steady state. For SFR$=3.8 M_{\odot}$~yr$^{-1}$ the mass-loading factor is then $\beta \equiv \dot{M}_{\text{outflow}}/\text{SFR} \sim 3$. This is similar to a powerful nuclear starburst \citep{bolatto13} or a quasar. 

In contrast, superbubble-driven winds have $\beta \ll 1$. Even if the models are badly wrong, there is a geometric issue: there appears to be more diffuse extraplanar gas than cool gas at the edges of chimneys \citep{howk00}, but there must be more gas to the sides of a bubble growing in an exponential disk than gas above or below it. This limits $\beta < 1$ for an individual superbubble, and likely much less. Finally, we expect most of the gas in superbubbles to be ejected hot \citep{weaver77,maclow89,roy13}, so the cool halo gas must condense out of it in this model. The cooling time is about $t_{\text{cool}} \sim 1.1 n_{-3}$~Gyr, where $n_{-3}$ is the density in units of $10^{-3}$~cm$^{-3}$ \citep[based on X-ray cooling curve at Solar metallicity;][]{schure09}. If we take an extreme value of $n_{-3} = 10$, the cooling time is comparable to the the orbital time, so the gas cannot cool fast enough in a classical fountain model. 

There are other models that can expel a lot of cold gas without expelling much hot gas, such as cosmic-ray driven winds \citep{breitschwerdt91} or radiation pressure on dust grains. Numerical simulations indicate that $\beta$ can be as high as $\sim$0.1-1 for cosmic-ray driven winds in $L_*$ disk galaxies \citep{booth13,salem14,ruszkowski17}, but this remains controversial \citep[e.g., $\beta\sim 0.001$ is reported by][]{fujita18}. $\beta > 1$ is expected primarily for smaller systems \citep{uhlig12}. Meanwhile, radiation pressure can only dominate the outflow rate when the dusty gas is near the Eddington limit, which would require SFR more than an order of magnitude greater for NGC~891 \citep{zhang_dong12,chattopadhyay12,krumholz12,zhang_dong17}. Neither of these models is a good candidate to explain the balance of hot and cool gas around NGC~891, especially the presence of molecular gas. It is probably possible to modify either model to increase $\beta$, but doing so could easily result in a model that is inconsistent with lower mass galaxies, as these winds should be stronger in smaller potentials. 

Alternatively, the gas may have been thrown out of the galaxy by a burst of past activity. Because the cool gas is not buoyant and will fall back within a few hundred Myr, it must have been expelled cold about this long ago or recently cooled from the hot gas. In the latter case, this would imply an extreme 0.3-10~keV $L_X > 10^{41}$~erg~s$^{-1}$ within the past few hundred Myr. This level of cooling is only seen in luminous infrared galaxies, and we would expect to see the effects of a major recent starburst in other bands. Instead, NGC~891 appears to be a normal galaxy.

We have found no satisfactory explanation for either the $M_{\text{cold}}/M_{\text{hot}} > 10$ ratio or the $(M_{\text{cold}}/t_{\text{orb}})/\text{SFR} \sim 3$ ratio. This suggests that the halo is not in a steady state, but there is no clear evidence that the SFR or AGN activity was much higher in the recent past. The large $M_{\text{cold}}/M_{\text{hot}}$ and fallback time of the cold gas also imply that the cool gas is mostly disconnected from the hot gas. At least in the steady state, this poses a challenge to stellar feedback models.

\subsection{Nonthermal Emission}

The \XMM{} and \Chandra{} 2-5~keV images reveal a diffuse, nonthermal glow that is concentrated around the galactic center, extends to a height of 5-6~kpc above the disk, and is consistent with the $\alpha=0.4\pm0.1$ power-law component measured in the halo spectra. The nonthermal emission could either be photons boosted by inverse Compton scattering off cosmic rays or synchrotron emission from high-energy cosmic rays, and the morphology suggests that they are associated with an ongoing or recent outburst from the galactic center.

This glow has a luminosity of $L_X \sim 3\times 10^{39}$~erg~s$^{-1}$, which is about ten times higher than that of the weakly accreting supermassive black hole \citep[$L_X \sim 3\times 10^{38}$~erg~s$^{-1}$;][]{she17}. If the nonthermal emission is from an AGN outburst, it must be a relic. In contrast, the SNe energy is sufficient (requiring one SN per $10^4$~yr), although it is not clear how SNe would produce this nonthermal X-ray emission. 

Inverse Compton scattering can be ruled out by the morphology, spectral index, and luminosity of the low frequency radio emission, as well as the magnetic field strength in the halo. \citet{mulcahy18} presented a deep 146~MHz map of the radio halo, where they see a thick disk-like morphology with no hint of the bubble-like hard X-ray emission. In addition, they measured a 146~MHz--5~GHz spectral index of $0.5 < \alpha < 0.7$ in the halo, whereas we measured $\alpha_X = 0.4\pm0.1$ even far from the disk. This could indicate that the up-scattered photons are from even lower energy electrons than those that radiate at 146~MHz, but this is unlikely because spectral ageing has yet to affect these electrons. Moreover, \citet{mulcahy18} and \citet{dahlem94} measured a halo $B$-field strength of $\approx 7$~$\mu$G, for which the synchrotron critical frequency of 146~MHz corresponds to a Lorentz factor of $\gamma \sim 2000$. Such electrons will up-scatter the cosmic microwave background (which peaks at $\nu = 160$~GHz) to $E \sim 3.5$~keV. This process is inevitable, but if CMB photons are the main inverse Compton coolant, the ratio of inverse Compton to synchrotron power is $u_{\text{rad}}/u_B \approx 0.2$. The 146~MHz luminosity $\nu L_{\nu} \sim 5\times 10^{36}$~erg~s$^{-1}$ predicts a hard X-ray luminosity of $L_X \sim 10^{36}$~erg~s$^{-1}$, which is more than two orders of magnitude below the observed value in the 2-5~keV band. 

We can also rule out the possibility that inverse Compton scattering of lower energy electrons ($\gamma < 1000$) by higher energy photons (the infrared interstellar radiation field) can explain the X-ray emission. Instead, scattering from the cosmic microwave background dominates in ``normal'' star-forming galaxies \citep[SFR=$2 M_{\odot}$~yr$^{-1}$;][]{schober15}, and the situation becomes worse off-plane. SFR$\gtrsim 10 M_{\odot}$~yr$^{-1}$ is needed for infrared photons to compete with the CMB a few kpc off-plane. 

Radio data at higher frequencies also rule out the X-rays as the high-energy tail of the synchrotron emission from a single electron population. The 1.4-5~GHz spectral index is $\alpha > 1.0$ in the halo \citep{wiegert15}, so $\alpha$ would have to be at least this steep. Instead, we measured $\alpha_X = 0.4\pm0.1$, which indicates that the X-ray emitting cosmic rays have yet to lose much energy. Finally, the total radio luminosity up to 5~GHz in the vicinity of the hard X-ray emission is a few$\times 10^{37}$~erg~s$^{-1}$, which is well below the X-ray luminosity of a few$\times 10^{39}$~erg~s$^{-1}$.

This leaves the possibility of a second population of high energy electrons which emit synchrotron emission in the X-ray band. For $B=7$~$\mu$G and assuming the cosmic rays are electrons, this would imply $\gamma > 10^8$. As far as we know, NGC~891 is not a $\gamma$-ray source, but it is near 3C~66A which is a bright source, so we cannot rule it out. A power-law index $\alpha_X = 0.4$ and a 0.01-10~keV $L_X \sim 3\times 10^{39}$~erg~s$^{-1}$ implies a $0.1-1$~MeV $L \sim 4\times 10^{40}$~erg~s$^{-1}$, which is likely too low to detect with any planned missions. Existing instruments are too insensitive and have poor resolution. We note that this scenario faces the difficulty that the nonthermal X-ray luminosity is high compared to the current SFR. We processed \XMM{} data for several other edge-on galaxies with higher SFR, and found no extraplanar, diffuse hard X-rays above about 2~kpc. 

None of these explanations seems likely, but a re-examination of the magnetic field strength or structure near the gaalctic center may provide stronger insights into the origin of this emission. 

\subsection{The 2T Model}

Previous authors have found that a 2T model adequately describes coronal spectra from most highly inclined disk galaxies \citep{dahlem98,tullmann06,li13a}, and \citet{strickland04} even found that a sample of galaxies can be jointly fitted by the \textit{same} 2T model. However, \citet{kuntz10} showed that, at low $S/N$, an arbitrarily complex temperature distribution can be fit by a 2T model and that the temperatures do not cleanly map onto the input distribution. Moreover, \citet{li13a} found $kT_1 = 0.2-0.3$~keV and $kT_2 = 0.4-1.0$~keV in a \Chandra{} sample of $L_*$ galaxies, whereas \citet{tullmann06} found that $kT_1 = 0.05-0.15$~keV and $kT_2 = 0.2-0.3$~keV from a \XMM{} sample that included many of the same galaxies. Finally, normal and starburst galaxies often have similar $kT_1$ and $kT_2$ \citep{strickland04,li13a}, even though the starbursts have brighter halos and different morphologies. 

The 2T model is a good fit to the halo spectrum of NGC~891 (Table~\ref{table.spectra}), and the temperature map reveals that the cooler and hotter gas have different spatial distributions (Figure~\ref{figure.tempmap}). Our spectral analysis further showed that the hot gas is concentrated in the inner halo and near the filament, so we propose that the 2T description from previous work is essentially accurate, with the cooler component corresponding to an extended (or at least relaxed) hot halo. Indeed, 2T models typically find one dominant component and one weaker component, with the dominant temperature close to the virial temperature for galaxies at or above $L_*$ \citep{li17}. 

The \citet{tullmann06} 2T models (based on the pn spectra) may not be accurate because they did not account for residual soft proton flares, which are strongest in the pn. Since flaring is concentrated near the optical axis, on-field background subtraction will underestimate this component, which has a soft, unabsorbed spectrum. This can be fit by an isothermal model with very low metallicity and zero absorption, or a 2T model with a cool component. \citet{tullmann06} account for the Galactic $N_{\text{H}}$, so a 2T model with $kT_1 < 0.15$~keV is needed, and the dominant temperature is slightly higher than $kT_{\text{vir}}$ to account for the presence of some hotter gas amidst the extended hot halo. We extracted spectra from several of the data sets they used, and we find that a hotter 2T model is a better description. 

\subsection{Metallicity and Other Galaxies}

Early measurements of the metallicity in galaxy groups and clusters underestimated the metallicity when fitting isothermal models to low-resolution spectra. At higher $S/N$ or spectral resolution, it became apparent that the hot gas is multi-phase and is highly enriched. Since the metallicity was measured from Fe lines, this became known as the ``iron bias'' \citep{buote98,buote00}. A similar problem can occur in galaxies, where \ion{O}{7}, \ion{O}{8}, and \ion{Fe}{17} produce the strongest lines. In addition, for the lower-temperature halos around galaxies, the metallicity measurement is susceptible to an incorrect model of the low-energy pseudo-continuum. For example, residual proton flares in the pn spectrum can lead to $Z/Z_{\odot} < 0.2$ regardless of the true value. In general, the non-X-ray background for each instrument has a soft energy spectrum and must be carefully modeled or subtracted, as 10\% errors severely bias the metallicity. 

The data presented here provide the best opportunity to measure the metallicity for an external $L_*$ galaxy halo, and our measurement of $Z/Z_{\odot}=0.14$ lends credence to measurements of $Z/Z_{\odot} \sim 0.1-0.2$ made at lower $S/N$ in the extended halos of other galaxies \citep{anderson11,bogdan17,bregman18}. If the bias in these galaxies is similar to that in NGC~891, the overall interpretation will not be thrown into doubt. If NGC~891 retains a virialized, low $Z/Z_{\odot}$ halo, then we would expect this situation in other late-type galaxies. 

The main counterpoint appears to be the Milky Way itself, for which \citet{bregman18} find $Z/Z_{\odot} \sim 0.5$ by using the pulsar dispersion measure to the Large Magellanic Cloud along with the empirical density and temperature profiles. Since NGC~891 is forming more stars, and has a larger \ion{H}{1} halo than the Milky Way, this result is surprising. However, the average metallicity measured towards the LMC out of the disk can be biased by some highly enriched material close to the disk and along the line of sight; decomposing the line profile with a high-resolution grating spectrometer such as \textit{Arcus} \citep{smith17} is necessary to determine this.

Finally, we note that X-ray faint elliptical galaxies demonstrate that the gas-phase metallicity can be low without resorting to dilution by pristine gas. These galaxies tend to have $Z/Z_{\odot} < 0.2$, but \citet{su13} ruled out accretion as a reason. The remaining possibilities are condensation of SNe ejecta into warm gas or Fe-rich dust or segregation into a hot wind \citep[][and references therein]{pipino11}. As we argued above, in star-forming galaxies (SFR$\gtrsim 1 M_{\odot}$~yr$^{-1}$) SN~II will drive feedback, and this makes it harder to segregate the metals. Thus, our assertion that other measurements of low metallicity in lower signal data sets imply that $L_*$ galaxies have extended, accreted halos applies only to the late-type galaxies. 

\section{Summary and Conclusions}
\label{section.summary}

We obtained a 320~ks \XMM{} observation of NGC~891. Combined with archival \XMM{} and \Chandra{} data, as well as data at other wavelengths, we found that:
\begin{itemize}
\item The hot halo mass and luminosity is dominated by gas at $kT=0.20\pm0.01$~keV with $Z/Z_{\odot}=0.14\pm0.03$(stat)$^{+0.08}_{-0.02}$(sys). A different choice of thermal model or Solar abundance ratio table than we adopted can lead to additional shifts of $\Delta Z/Z_{\odot} < \pm0.05$ for commonly used, published models. This gas accounts for most of the emission even near the disk. We examined issues with measuring the metallicity at low spectral resolution, and found that the choice of model and Solar abundance table makes about as much difference as the possible systematic bias. This gas surrounds the optical disk, and there is a one-sided filament above the galactic center that is visible up to 15~kpc. The luminosity implies a cooling rate of $\dot{M} \sim 0.5 M_{\odot}$~yr$^{-1}$, which is the accretion rate onto the disk in the steady state. 
\item In addition to the virialized hot gas, there is hotter gas with $kT=0.71\pm0.04$~keV that is concentrated above star-forming regions in the disk and in the filament. An isothermal temperature map shows that this gas has a substantially different spatial distribution than the virialized gas, so it is likely an outflow.  Its metallicity is indeterminate.
\item There is diffuse, non-thermal, hard X-ray emission concentrated around the galactic center that is visible up to 5-6~kpc above the disk and which has a spectral index $\alpha=0.4\pm0.1$. This cannot be explained by instrumental backgrounds or astrophysical foregrounds, and is likely associated with NGC~891. With a luminosity comparable to that of the hot gas, the non-thermal emission is too bright to arise from ongoing AGN or SNe activity, but may be a relic. A comparison with the radio data indicates that it cannot be inverse-Compton scattering from the CMB, nor can it be the high energy tail of the radio synchrotron emission. It is possible that it is the low energy part of a second synchrotron population.
\item The accuracy of the metallicity measurement cannot be improved at the same spectral resolution by increasing the $S/N$. Spectrometers with higher energy resolution are needed to reduce systematic error and distinguish between plasma emission models; the \textit{Hitomi} Soft X-ray Spectrometer measurements of the Perseus Cluster \citep{hitomi16} have already shown that updates to these models are needed.
\end{itemize}
The low metallicity in the X-ray emitting gas is consistent with the visible corona being the innermost part of an extended hot halo accreted from the intergalactic medium. The morphology of the halo near the disk will be affected by feedback activity, and this may explain the ubiquity of X-ray coronae that appear to be thick disks near the virial temperature. However, further numerical work is needed to explore the interaction of enriched outflows with a low metallicity halo. 

The large disparity between the amount of hot and cold gas in the halo challenge models in which SNe activity ejects hot gas that cools in the halo and falls back. Any such model would need extreme mass-loading factors, and again this calls for further theoretical work to demonstrate feasibility. 

These measurements, and especially spatially resolving the different temperature components, indicate that the 2T model that is generally successful for halo spectra \citep{strickland04,li13a} is essentially correct: there is a virialized hot halo (which could have either solar or sub-solar metallicity depending on its origin) that co-exists with ongoing, SNe-powered outflows. Moreover, these measurements indicate that measurements of the metallicity around other galaxies made at lower $S/N$ are basically correct. At least for late-type galaxies, galactic coronae appear to be consistent with the $\Lambda$CDM picture of a hot halo built up from cosmic infall.

\acknowledgments
The authors thank the anonymous referee for a detailed report that improved the clarity and content of this manuscript. Support for this work was provided by the National Aeronautics and Space Administration through \XMM{} Award Number 078076 issued by the \XMM{} Guest Observer Facility, under contract NNX17AD57G.

This research has made use of the NASA/IPAC Extragalactic Database (NED) which is operated by the Jet Propulsion Laboratory, California Institute of Technology, under contract with the National Aeronautics and Space Administration. We acknowledge the usage of the HyperLeda database (http://leda.univ-lyon1.fr). The scientific results in this article are based in part on observations made by the \Chandra{} X-ray Observatory and published previously, as well as observations performed by the Karl~G.~Jansky Very Large Array. The National Radio Astronomy Observatory is a facility of the National Science Foundation operated under cooperative agreement by Associated Universities, Inc.


\begin{thebibliography}{121}
\expandafter\ifx\csname natexlab\endcsname\relax\def\natexlab#1{#1}\fi

\bibitem[{{Aguirre} {et~al.}(2008){Aguirre}, {Dow-Hygelund}, {Schaye}, \&
  {Theuns}}]{aguirre08}
{Aguirre}, A., {Dow-Hygelund}, C., {Schaye}, J., \& {Theuns}, T. 2008, \apj,
  689, 851

\bibitem[{{Allen} {et~al.}(1978){Allen}, {Baldwin}, \& {Sancisi}}]{allen78}
{Allen}, R.~J., {Baldwin}, J.~E., \& {Sancisi}, R. 1978, \aap, 62, 397

\bibitem[{{Anders} \& {Grevesse}(1989)}]{anders89}
{Anders}, E., \& {Grevesse}, N. 1989, \gca, 53, 197

\bibitem[{{Anderson} \& {Bregman}(2011)}]{anderson11}
{Anderson}, M.~E., \& {Bregman}, J.~N. 2011, \apj, 737, 22

\bibitem[{{Anderson} {et~al.}(2013){Anderson}, {Bregman}, \&
  {Dai}}]{anderson13}
{Anderson}, M.~E., {Bregman}, J.~N., \& {Dai}, X. 2013, \apj, 762, 106

\bibitem[{{Anderson} {et~al.}(2016){Anderson}, {Churazov}, \&
  {Bregman}}]{anderson16}
{Anderson}, M.~E., {Churazov}, E., \& {Bregman}, J.~N. 2016, \mnras, 455, 227

\bibitem[{{Armillotta} {et~al.}(2016){Armillotta}, {Fraternali}, \&
  {Marinacci}}]{armillotta16}
{Armillotta}, L., {Fraternali}, F., \& {Marinacci}, F. 2016, \mnras, 462, 4157

\bibitem[{{Arnaud}(1996)}]{arnaud96}
{Arnaud}, K.~A. 1996, in Astronomical Society of the Pacific Conference Series,
  Vol. 101, Astronomical Data Analysis Software and Systems V, ed.
  {G.~H.~Jacoby \& J.~Barnes}, 17

\bibitem[{{Asplund} {et~al.}(2009){Asplund}, {Grevesse}, {Sauval}, \&
  {Scott}}]{asplund09}
{Asplund}, M., {Grevesse}, N., {Sauval}, A.~J., \& {Scott}, P. 2009, \araa, 47,
  481

\bibitem[{{Baumgartner} \& {Breitschwerdt}(2013)}]{baumgartner13}
{Baumgartner}, V., \& {Breitschwerdt}, D. 2013, \aap, 557, A140

\bibitem[{{Bocchio} {et~al.}(2016){Bocchio}, {Bianchi}, {Hunt}, \&
  {Schneider}}]{bocchio16}
{Bocchio}, M., {Bianchi}, S., {Hunt}, L.~K., \& {Schneider}, R. 2016, \aap,
  586, A8

\bibitem[{{Boettcher} {et~al.}(2016){Boettcher}, {Zweibel}, {Gallagher}, \&
  {Benjamin}}]{boettcher16}
{Boettcher}, E., {Zweibel}, E.~G., {Gallagher}, III, J.~S., \& {Benjamin},
  R.~A. 2016, \apj, 832, 118

\bibitem[{{Bogd{\'a}n} {et~al.}(2017){Bogd{\'a}n}, {Bourdin}, {Forman},
  {Kraft}, {Vogelsberger}, {Hernquist}, \& {Springel}}]{bogdan17}
{Bogd{\'a}n}, {\'A}., {Bourdin}, H., {Forman}, W.~R., {Kraft}, R.~P.,
  {Vogelsberger}, M., {Hernquist}, L., \& {Springel}, V. 2017, \apj, 850, 98

\bibitem[{{Bogd{\'a}n} {et~al.}(2013){Bogd{\'a}n}, {Forman}, {Kraft}, \&
  {Jones}}]{bogdan13}
{Bogd{\'a}n}, {\'A}., {Forman}, W.~R., {Kraft}, R.~P., \& {Jones}, C. 2013,
  \apj, 772, 98

\bibitem[{{Bolatto} {et~al.}(2013){Bolatto}, {Warren}, {Leroy}, {Walter},
  {Veilleux}, {Ostriker}, {Ott}, {Zwaan}, {Fisher}, {Weiss}, {Rosolowsky}, \&
  {Hodge}}]{bolatto13}
{Bolatto}, A.~D., {et~al.} 2013, \nat, 499, 450

\bibitem[{{Booth} {et~al.}(2013){Booth}, {Agertz}, {Kravtsov}, \&
  {Gnedin}}]{booth13}
{Booth}, C.~M., {Agertz}, O., {Kravtsov}, A.~V., \& {Gnedin}, N.~Y. 2013,
  \apjl, 777, L16

\bibitem[{{Bregman} {et~al.}(2018){Bregman}, {Anderson}, {Miller},
  {Hodges-Kluck}, {Dai}, {Li}, {Li}, \& {Qu}}]{bregman18}
{Bregman}, J.~N., {Anderson}, M.~E., {Miller}, M.~J., {Hodges-Kluck}, E.,
  {Dai}, X., {Li}, J.-T., {Li}, Y., \& {Qu}, Z. 2018, ArXiv e-prints

\bibitem[{{Bregman} \& {Pildis}(1994)}]{bregman94}
{Bregman}, J.~N., \& {Pildis}, R.~A. 1994, \apj, 420, 570

\bibitem[{{Breitschwerdt} {et~al.}(1991){Breitschwerdt}, {McKenzie}, \&
  {Voelk}}]{breitschwerdt91}
{Breitschwerdt}, D., {McKenzie}, J.~F., \& {Voelk}, H.~J. 1991, \aap, 245, 79

\bibitem[{{Bresolin}(2007)}]{bresolin07}
{Bresolin}, F. 2007, \apj, 656, 186

\bibitem[{{Buote}(2000)}]{buote00}
{Buote}, D.~A. 2000, \mnras, 311, 176

\bibitem[{{Buote} \& {Fabian}(1998)}]{buote98}
{Buote}, D.~A., \& {Fabian}, A.~C. 1998, \mnras, 296, 977

\bibitem[{{Carter} \& {Sembay}(2008)}]{carter08}
{Carter}, J.~A., \& {Sembay}, S. 2008, \aap, 489, 837

\bibitem[{{Cash}(1979)}]{cash79}
{Cash}, W. 1979, \apj, 228, 939

\bibitem[{{Cavaliere} \& {Fusco-Femiano}(1976)}]{cavaliere76}
{Cavaliere}, A., \& {Fusco-Femiano}, R. 1976, \aap, 49, 137

\bibitem[{{Chattopadhyay} {et~al.}(2012){Chattopadhyay}, {Sharma}, {Nath}, \&
  {Ryu}}]{chattopadhyay12}
{Chattopadhyay}, I., {Sharma}, M., {Nath}, B.~B., \& {Ryu}, D. 2012, \mnras,
  423, 2153

\bibitem[{{Crain} {et~al.}(2010){Crain}, {McCarthy}, {Frenk}, {Theuns}, \&
  {Schaye}}]{crain10}
{Crain}, R.~A., {McCarthy}, I.~G., {Frenk}, C.~S., {Theuns}, T., \& {Schaye},
  J. 2010, \mnras, 407, 1403

\bibitem[{{Dahlem} {et~al.}(1994){Dahlem}, {Dettmar}, \& {Hummel}}]{dahlem94}
{Dahlem}, M., {Dettmar}, R.-J., \& {Hummel}, E. 1994, \aap, 290, 384

\bibitem[{{Dahlem} {et~al.}(1998){Dahlem}, {Weaver}, \& {Heckman}}]{dahlem98}
{Dahlem}, M., {Weaver}, K.~A., \& {Heckman}, T.~M. 1998, \apjs, 118, 401

\bibitem[{{Dai} {et~al.}(2012){Dai}, {Anderson}, {Bregman}, \&
  {Miller}}]{dai12}
{Dai}, X., {Anderson}, M.~E., {Bregman}, J.~N., \& {Miller}, J.~M. 2012, \apj,
  755, 107

\bibitem[{{de Vaucouleurs} {et~al.}(1991){de Vaucouleurs}, {de Vaucouleurs},
  {Corwin}, {Buta}, {Paturel}, \& {Fouque}}]{devaucouleurs91}
{de Vaucouleurs}, G., {de Vaucouleurs}, A., {Corwin}, Jr., H.~G., {Buta},
  R.~J., {Paturel}, G., \& {Fouque}, P. 1991, {Third Reference Catalogue of
  Bright Galaxies}, ed. {Roman, N.~G., de Vaucouleurs, G., de Vaucouleurs, A.,
  Corwin, H.~G., Jr., Buta, R.~J., Paturel, G., \& Fouqu{\'e}, P.}

\bibitem[{{Dickey} \& {Lockman}(1990)}]{dickey90}
{Dickey}, J.~M., \& {Lockman}, F.~J. 1990, \araa, 28, 215

\bibitem[{{Faerman} {et~al.}(2017){Faerman}, {Sternberg}, \&
  {McKee}}]{faerman17}
{Faerman}, Y., {Sternberg}, A., \& {McKee}, C.~F. 2017, \apj, 835, 52

\bibitem[{{Fioretti} {et~al.}(2016){Fioretti}, {Bulgarelli}, {Malaguti},
  {Spiga}, \& {Tiengo}}]{fioretti16}
{Fioretti}, V., {Bulgarelli}, A., {Malaguti}, G., {Spiga}, D., \& {Tiengo}, A.
  2016, in \procspie, Vol. 9905, Space Telescopes and Instrumentation 2016:
  Ultraviolet to Gamma Ray, 99056W

\bibitem[{{Foster} {et~al.}(2012){Foster}, {Ji}, {Smith}, \&
  {Brickhouse}}]{foster12}
{Foster}, A.~R., {Ji}, L., {Smith}, R.~K., \& {Brickhouse}, N.~S. 2012, \apj,
  756, 128

\bibitem[{{Fraternali}(2017)}]{fraternali17}
{Fraternali}, F. 2017, in Astrophysics and Space Science Library, Vol. 430, Gas
  Accretion onto Galaxies, ed. A.~{Fox} \& R.~{Dav{\'e}}, 323

\bibitem[{{Fraternali} \& {Binney}(2006)}]{fraternali06}
{Fraternali}, F., \& {Binney}, J.~J. 2006, \mnras, 366, 449

\bibitem[{{Fujita} \& {Mac Low}(2018)}]{fujita18}
{Fujita}, A., \& {Mac Low}, M.-M. 2018, \mnras, 477, 531

\bibitem[{{Garcia-Burillo} \& {Guelin}(1995)}]{garcia-burillo95}
{Garcia-Burillo}, S., \& {Guelin}, M. 1995, \aap, 299, 657

\bibitem[{{Garcia-Burillo} {et~al.}(1992){Garcia-Burillo}, {Guelin},
  {Cernicharo}, \& {Dahlem}}]{garcia-burillo92}
{Garcia-Burillo}, S., {Guelin}, M., {Cernicharo}, J., \& {Dahlem}, M. 1992,
  \aap, 266, 21

\bibitem[{{Gaskin} {et~al.}(2018){Gaskin}, {Dominguez}, {Gelmis}, {Mulqueen},
  {Swartz}, {McCarley}, {{\"O}zel}, {Vikhlinin}, {Schwartz}, {Tananbaum},
  {Blackwood}, {Arenberg}, {Purcell}, \& {Allen}}]{gaskin18}
{Gaskin}, J.~A., {et~al.} 2018, in Society of Photo-Optical Instrumentation
  Engineers (SPIE) Conference Series, Vol. 10699, 106990N

\bibitem[{{Gastaldello} {et~al.}(2017){Gastaldello}, {Ghizzardi}, {Marelli},
  {Salvetti}, {Molendi}, {De Luca}, {Moretti}, {Rossetti}, \&
  {Tiengo}}]{gastaldello17}
{Gastaldello}, F., {et~al.} 2017, Experimental Astronomy, 44, 321

\bibitem[{{Gilfanov}(2004)}]{gilfanov04}
{Gilfanov}, M. 2004, \mnras, 349, 146

\bibitem[{{Grevesse} \& {Sauval}(1998)}]{grevesse98}
{Grevesse}, N., \& {Sauval}, A.~J. 1998, \ssr, 85, 161

\bibitem[{{Gupta} {et~al.}(2014){Gupta}, {Mathur}, {Galeazzi}, \&
  {Krongold}}]{gupta14}
{Gupta}, A., {Mathur}, S., {Galeazzi}, M., \& {Krongold}, Y. 2014, \apss, 352,
  775

\bibitem[{{Heckman} \& {Thompson}(2017)}]{heckman17}
{Heckman}, T.~M., \& {Thompson}, T.~A. 2017, ArXiv e-prints

\bibitem[{{Henley} \& {Shelton}(2013)}]{henley13}
{Henley}, D.~B., \& {Shelton}, R.~L. 2013, \apj, 773, 92

\bibitem[{{Hitomi Collaboration} {et~al.}(2016){Hitomi Collaboration},
  {Aharonian}, {Akamatsu}, {Akimoto}, {Allen}, {Anabuki}, {Angelini}, {Arnaud},
  {Audard}, {Awaki}, {Axelsson}, {Bamba}, {Bautz}, {Blandford}, {Brenneman},
  {Brown}, {Bulbul}, {Cackett}, {Chernyakova}, {Chiao}, {Coppi}, {Costantini},
  {de Plaa}, {den Herder}, {Done}, {Dotani}, {Ebisawa}, {Eckart}, {Enoto},
  {Ezoe}, {Fabian}, {Ferrigno}, {Foster}, {Fujimoto}, {Fukazawa}, {Furuzawa},
  {Galeazzi}, {Gallo}, {Gandhi}, {Giustini}, {Goldwurm}, {Gu}, {Guainazzi},
  {Haba}, {Hagino}, {Hamaguchi}, {Harrus}, {Hatsukade}, {Hayashi}, {Hayashi},
  {Hayashida}, {Hiraga}, {Hornschemeier}, {Hoshino}, {Hughes}, {Iizuka},
  {Inoue}, {Inoue}, {Ishibashi}, {Ishida}, {Ishikawa}, {Ishisaki}, {Itoh},
  {Iyomoto}, {Kaastra}, {Kallman}, {Kamae}, {Kara}, {Kataoka}, {Katsuda},
  {Katsuta}, {Kawaharada}, {Kawai}, {Kelley}, {Khangulyan}, {Kilbourne},
  {King}, {Kitaguchi}, {Kitamoto}, {Kitayama}, {Kohmura}, {Kokubun}, {Koyama},
  {Koyama}, {Kretschmar}, {Krimm}, {Kubota}, {Kunieda}, {Laurent}, {Lebrun},
  {Lee}, {Leutenegger}, {Limousin}, {Loewenstein}, {Long}, {Lumb}, {Madejski},
  {Maeda}, {Maier}, {Makishima}, {Markevitch}, {Matsumoto}, {Matsushita},
  {McCammon}, {McNamara}, {Mehdipour}, {Miller}, {Miller}, {Mineshige},
  {Mitsuda}, {Mitsuishi}, {Miyazawa}, {Mizuno}, {Mori}, {Mori}, {Moseley},
  {Mukai}, {Murakami}, {Murakami}, {Mushotzky}, {Nagino}, {Nakagawa},
  {Nakajima}, {Nakamori}, {Nakano}, {Nakashima}, {Nakazawa}, {Nobukawa},
  {Noda}, {Nomachi}, {O'Dell}, {Odaka}, {Ohashi}, {Ohno}, {Okajima}, {Ota},
  {Ozaki}, {Paerels}, {Paltani}, {Parmar}, {Petre}, {Pinto}, {Pohl}, {Porter},
  {Pottschmidt}, {Ramsey}, {Reynolds}, {Russell}, {Safi-Harb}, {Saito},
  {Sakai}, {Sameshima}, {Sato}, {Sato}, {Sato}, {Sawada}, {Schartel},
  {Serlemitsos}, {Seta}, {Shidatsu}, {Simionescu}, {Smith}, {Soong}, {Stawarz},
  {Sugawara}, {Sugita}, {Szymkowiak}, {Tajima}, {Takahashi}, {Takahashi},
  {Takeda}, {Takei}, {Tamagawa}, {Tamura}, {Tamura}, {Tanaka}, {Tanaka},
  {Tanaka}, {Tashiro}, {Tawara}, {Terada}, {Terashima}, {Tombesi}, {Tomida},
  {Tsuboi}, {Tsujimoto}, {Tsunemi}, {Tsuru}, {Uchida}, {Uchiyama}, {Uchiyama},
  {Ueda}, {Ueda}, {Ueno}, {Uno}, {Urry}, {Ursino}, {de Vries}, {Watanabe},
  {Werner}, {Wik}, {Wilkins}, {Williams}, {Yamada}, {Yamaguchi}, {Yamaoka},
  {Yamasaki}, {Yamauchi}, {Yamauchi}, {Yaqoob}, {Yatsu}, {Yonetoku}, {Yoshida},
  {Yuasa}, {Zhuravleva}, \& {Zoghbi}}]{hitomi16}
{Hitomi Collaboration} {et~al.} 2016, \nat, 535, 117

\bibitem[{{Hodges-Kluck} {et~al.}(2016){Hodges-Kluck}, {Cafmeyer}, \&
  {Bregman}}]{hk16}
{Hodges-Kluck}, E., {Cafmeyer}, J., \& {Bregman}, J.~N. 2016, \apj, 833, 58

\bibitem[{{Hodges-Kluck} \& {Bregman}(2013)}]{hk13}
{Hodges-Kluck}, E.~J., \& {Bregman}, J.~N. 2013, \apj, 762, 12

\bibitem[{{Hodges-Kluck} {et~al.}(2012){Hodges-Kluck}, {Bregman}, {Miller}, \&
  {Pellegrini}}]{hk12}
{Hodges-Kluck}, E.~J., {Bregman}, J.~N., {Miller}, J.~M., \& {Pellegrini}, E.
  2012, \apjl, 747, L39

\bibitem[{{Howk} \& {Savage}(2000)}]{howk00}
{Howk}, J.~C., \& {Savage}, B.~D. 2000, \aj, 119, 644

\bibitem[{{Hummel} {et~al.}(1991){Hummel}, {Dahlem}, {van der Hulst}, \&
  {Sukumar}}]{hummel91}
{Hummel}, E., {Dahlem}, M., {van der Hulst}, J.~M., \& {Sukumar}, S. 1991,
  \aap, 246, 10

\bibitem[{{Humphrey} \& {Buote}(2006)}]{humphrey06}
{Humphrey}, P.~J., \& {Buote}, D.~A. 2006, \apj, 639, 136

\bibitem[{{Humphrey} {et~al.}(2011){Humphrey}, {Buote}, {Canizares}, {Fabian},
  \& {Miller}}]{humphrey11}
{Humphrey}, P.~J., {Buote}, D.~A., {Canizares}, C.~R., {Fabian}, A.~C., \&
  {Miller}, J.~M. 2011, \apj, 729, 53

\bibitem[{{Irwin} {et~al.}(2012){Irwin}, {Beck}, {Benjamin}, {Dettmar},
  {English}, {Heald}, {Henriksen}, {Johnson}, {Krause}, {Li}, {Miskolczi},
  {Mora}, {Murphy}, {Oosterloo}, {Porter}, {Rand}, {Saikia}, {Schmidt},
  {Strong}, {Walterbos}, {Wang}, \& {Wiegert}}]{irwin12b}
{Irwin}, J., {et~al.} 2012, \aj, 144, 44

\bibitem[{{Kalberla} {et~al.}(2005){Kalberla}, {Burton}, {Hartmann}, {Arnal},
  {Bajaja}, {Morras}, \& {P{\"o}ppel}}]{kalberla05}
{Kalberla}, P.~M.~W., {Burton}, W.~B., {Hartmann}, D., {Arnal}, E.~M.,
  {Bajaja}, E., {Morras}, R., \& {P{\"o}ppel}, W.~G.~L. 2005, \aap, 440, 775

\bibitem[{{Kelley} {et~al.}(2016){Kelley}, {Akamatsu}, {Azzarello}, {Bialas},
  {Boyce}, {Brown}, {Canavan}, {Chiao}, {Costantini}, {DiPirro}, {Eckart},
  {Ezoe}, {Fujimoto}, {Haas}, {den Herder}, {Hoshino}, {Ishikawa}, {Ishisaki},
  {Iyomoto}, {Kilbourne}, {Kimball}, {Kitamoto}, {Konami}, {Koyama},
  {Leutenegger}, {McCammon}, {Mitsuda}, {Mitsuishi}, {Moseley}, {Murakami},
  {Murakami}, {Noda}, {Ogawa}, {Ohashi}, {Okamoto}, {Ota}, {Paltani}, {Porter},
  {Sakai}, {Sato}, {Sato}, {Sawada}, {Seta}, {Shinozaki}, {Shirron},
  {Sneiderman}, {Sugita}, {Szymkowiak}, {Takei}, {Tamagawa}, {Tashiro},
  {Terada}, {Tsujimoto}, {de Vries}, {Yamada}, {Yamasaki}, \&
  {Yatsu}}]{kelley16}
{Kelley}, R.~L., {et~al.} 2016, in \procspie, Vol. 9905, Space Telescopes and
  Instrumentation 2016: Ultraviolet to Gamma Ray, 99050V

\bibitem[{{Krause}(2009)}]{krause09}
{Krause}, M. 2009, in Revista Mexicana de Astronomia y Astrofisica Conference
  Series, Vol.~36, Revista Mexicana de Astronomia y Astrofisica Conference
  Series, 25--29

\bibitem[{{Kroupa}(2002)}]{kroupa02}
{Kroupa}, P. 2002, Science, 295, 82

\bibitem[{{Krumholz} \& {Thompson}(2012)}]{krumholz12}
{Krumholz}, M.~R., \& {Thompson}, T.~A. 2012, \apj, 760, 155

\bibitem[{{Kudritzki} {et~al.}(2015){Kudritzki}, {Ho}, {Schruba}, {Burkert},
  {Zahid}, {Bresolin}, \& {Dima}}]{kudritzki15}
{Kudritzki}, R.-P., {Ho}, I.-T., {Schruba}, A., {Burkert}, A., {Zahid}, H.~J.,
  {Bresolin}, F., \& {Dima}, G.~I. 2015, \mnras, 450, 342

\bibitem[{{Kuntz} \& {Snowden}(2010)}]{kuntz10}
{Kuntz}, K.~D., \& {Snowden}, S.~L. 2010, \apjs, 188, 46

\bibitem[{{Lehmer} {et~al.}(2010){Lehmer}, {Alexander}, {Bauer}, {Brandt},
  {Goulding}, {Jenkins}, {Ptak}, \& {Roberts}}]{lehmer10}
{Lehmer}, B.~D., {Alexander}, D.~M., {Bauer}, F.~E., {Brandt}, W.~N.,
  {Goulding}, A.~D., {Jenkins}, L.~P., {Ptak}, A., \& {Roberts}, T.~P. 2010,
  \apj, 724, 559

\bibitem[{{Li}(2015)}]{li15}
{Li}, J.-T. 2015, \mnras, 453, 1062

\bibitem[{{Li} {et~al.}(2017){Li}, {Bregman}, {Wang}, {Crain}, {Anderson}, \&
  {Zhang}}]{li17}
{Li}, J.-T., {Bregman}, J.~N., {Wang}, Q.~D., {Crain}, R.~A., {Anderson},
  M.~E., \& {Zhang}, S. 2017, \apjs, 233, 20

\bibitem[{{Li} {et~al.}(2008){Li}, {Li}, {Wang}, {Irwin}, \& {Rossa}}]{li08}
{Li}, J.-T., {Li}, Z., {Wang}, Q.~D., {Irwin}, J.~A., \& {Rossa}, J. 2008,
  \mnras, 390, 59

\bibitem[{{Li} \& {Wang}(2013)}]{li13a}
{Li}, J.-T., \& {Wang}, Q.~D. 2013, \mnras, 428, 2085

\bibitem[{{Li} \& {Bregman}(2017)}]{li_yunyang17}
{Li}, Y., \& {Bregman}, J. 2017, \apj, 849, 105

\bibitem[{{Li} {et~al.}(2011){Li}, {Jones}, {Forman}, {Kraft}, {Lal}, {Di
  Stefano}, {Spitler}, {Tang}, {Wang}, {Gilfanov}, \&
  {Revnivtsev}}]{li_zhiyuan11}
{Li}, Z., {et~al.} 2011, \apj, 730, 84

\bibitem[{{Liedahl} {et~al.}(1995){Liedahl}, {Osterheld}, \&
  {Goldstein}}]{liedahl95}
{Liedahl}, D.~A., {Osterheld}, A.~L., \& {Goldstein}, W.~H. 1995, \apjl, 438,
  L115

\bibitem[{{Mac Low} {et~al.}(1989){Mac Low}, {McCray}, \& {Norman}}]{maclow89}
{Mac Low}, M.-M., {McCray}, R., \& {Norman}, M.~L. 1989, \apj, 337, 141

\bibitem[{{Mannucci} {et~al.}(2005){Mannucci}, {Della Valle}, {Panagia},
  {Cappellaro}, {Cresci}, {Maiolino}, {Petrosian}, \& {Turatto}}]{mannucci05}
{Mannucci}, F., {Della Valle}, M., {Panagia}, N., {Cappellaro}, E., {Cresci},
  G., {Maiolino}, R., {Petrosian}, A., \& {Turatto}, M. 2005, \aap, 433, 807

\bibitem[{{Marinacci} {et~al.}(2010){Marinacci}, {Binney}, {Fraternali},
  {Nipoti}, {Ciotti}, \& {Londrillo}}]{marinacci10}
{Marinacci}, F., {Binney}, J., {Fraternali}, F., {Nipoti}, C., {Ciotti}, L., \&
  {Londrillo}, P. 2010, \mnras, 404, 1464

\bibitem[{{Melioli} {et~al.}(2009){Melioli}, {Brighenti}, {D'Ercole}, \& {de
  Gouveia Dal Pino}}]{melioli09}
{Melioli}, C., {Brighenti}, F., {D'Ercole}, A., \& {de Gouveia Dal Pino}, E.~M.
  2009, \mnras, 399, 1089

\bibitem[{{Mewe} {et~al.}(1985){Mewe}, {Gronenschild}, \& {van den
  Oord}}]{mewe85}
{Mewe}, R., {Gronenschild}, E.~H.~B.~M., \& {van den Oord}, G.~H.~J. 1985,
  \aaps, 62, 197

\bibitem[{{Mewe} {et~al.}(1986){Mewe}, {Lemen}, \& {van den Oord}}]{mewe86}
{Mewe}, R., {Lemen}, J.~R., \& {van den Oord}, G.~H.~J. 1986, \aaps, 65, 511

\bibitem[{{Miller} \& {Bregman}(2015)}]{miller15}
{Miller}, M.~J., \& {Bregman}, J.~N. 2015, \apj, 800, 14

\bibitem[{{Mineo} {et~al.}(2012){Mineo}, {Gilfanov}, \& {Sunyaev}}]{mineo12}
{Mineo}, S., {Gilfanov}, M., \& {Sunyaev}, R. 2012, \mnras, 419, 2095

\bibitem[{{Mineshige} {et~al.}(1994){Mineshige}, {Hirano}, {Kitamoto},
  {Yamada}, \& {Fukue}}]{mineshige94}
{Mineshige}, S., {Hirano}, A., {Kitamoto}, S., {Yamada}, T.~T., \& {Fukue}, J.
  1994, \apj, 426, 308

\bibitem[{{Mouhcine} {et~al.}(2010){Mouhcine}, {Ibata}, \&
  {Rejkuba}}]{mouhcine10}
{Mouhcine}, M., {Ibata}, R., \& {Rejkuba}, M. 2010, \apjl, 714, L12

\bibitem[{{Mulcahy} {et~al.}(2018){Mulcahy}, {Horneffer}, {Beck}, {Krause},
  {Schmidt}, {Basu}, {Chyzy}, {Dettmar}, {Haverkorn}, {Heald}, {Heesen},
  {Horellou}, {Iacobelli}, {Nikiel-Wroczynski}, {Paladino}, {Scaife},
  {Sridhar}, {Strom}, {Tabatabaei}, {Cantwel}, {Carey}, {Grainge}, {Hickish},
  {Perrot}, {Razavi-Ghods}, {Scott}, \& {Titterington}}]{mulcahy18}
{Mulcahy}, D.~D., {et~al.} 2018, ArXiv e-prints

\bibitem[{{Mushotzky}(2018)}]{mushotzky18}
{Mushotzky}, R. 2018, ArXiv e-prints

\bibitem[{{Nicastro} {et~al.}(2016){Nicastro}, {Senatore}, {Krongold},
  {Mathur}, \& {Elvis}}]{nicastro16}
{Nicastro}, F., {Senatore}, F., {Krongold}, Y., {Mathur}, S., \& {Elvis}, M.
  2016, \apjl, 828, L12

\bibitem[{{Nuza} {et~al.}(2014){Nuza}, {Parisi}, {Scannapieco}, {Richter},
  {Gottl{\"o}ber}, \& {Steinmetz}}]{nuza14}
{Nuza}, S.~E., {Parisi}, F., {Scannapieco}, C., {Richter}, P., {Gottl{\"o}ber},
  S., \& {Steinmetz}, M. 2014, \mnras, 441, 2593

\bibitem[{{Oosterloo} {et~al.}(2007){Oosterloo}, {Fraternali}, \&
  {Sancisi}}]{oosterloo07}
{Oosterloo}, T., {Fraternali}, F., \& {Sancisi}, R. 2007, \aj, 134, 1019

\bibitem[{{Pipino} \& {Matteucci}(2011)}]{pipino11}
{Pipino}, A., \& {Matteucci}, F. 2011, \aap, 530, A98

\bibitem[{{Popescu} {et~al.}(2004){Popescu}, {Tuffs}, {Kylafis}, \&
  {Madore}}]{popescu04}
{Popescu}, C.~C., {Tuffs}, R.~J., {Kylafis}, N.~D., \& {Madore}, B.~F. 2004,
  \aap, 414, 45

\bibitem[{{Rand}(1994)}]{rand94}
{Rand}, R.~J. 1994, \aap, 285, 833

\bibitem[{{Roy} {et~al.}(2013){Roy}, {Nath}, {Sharma}, \& {Shchekinov}}]{roy13}
{Roy}, A., {Nath}, B.~B., {Sharma}, P., \& {Shchekinov}, Y. 2013, \mnras, 434,
  3572

\bibitem[{{Rupen}(1991)}]{rupen91}
{Rupen}, M.~P. 1991, \aj, 102, 48

\bibitem[{{Ruszkowski} {et~al.}(2017){Ruszkowski}, {Yang}, \&
  {Zweibel}}]{ruszkowski17}
{Ruszkowski}, M., {Yang}, H.-Y.~K., \& {Zweibel}, E. 2017, \apj, 834, 208

\bibitem[{{Salem} \& {Bryan}(2014)}]{salem14}
{Salem}, M., \& {Bryan}, G.~L. 2014, \mnras, 437, 3312

\bibitem[{{Sanders} {et~al.}(2016){Sanders}, {Fabian}, {Russell}, {Walker}, \&
  {Blundell}}]{sanders16}
{Sanders}, J.~S., {Fabian}, A.~C., {Russell}, H.~R., {Walker}, S.~A., \&
  {Blundell}, K.~M. 2016, \mnras, 460, 1898

\bibitem[{{Schechter}(1976)}]{schechter76}
{Schechter}, P. 1976, \apj, 203, 297

\bibitem[{{Schober} {et~al.}(2015){Schober}, {Schleicher}, \&
  {Klessen}}]{schober15}
{Schober}, J., {Schleicher}, D.~R.~G., \& {Klessen}, R.~S. 2015, \mnras, 446, 2

\bibitem[{{Schure} {et~al.}(2009){Schure}, {Kosenko}, {Kaastra}, {Keppens}, \&
  {Vink}}]{schure09}
{Schure}, K.~M., {Kosenko}, D., {Kaastra}, J.~S., {Keppens}, R., \& {Vink}, J.
  2009, \aap, 508, 751

\bibitem[{{She} {et~al.}(2017){She}, {Ho}, \& {Feng}}]{she17}
{She}, R., {Ho}, L.~C., \& {Feng}, H. 2017, \apj, 842, 131

\bibitem[{{Shen} {et~al.}(2010){Shen}, {Wadsley}, \& {Stinson}}]{shen10}
{Shen}, S., {Wadsley}, J., \& {Stinson}, G. 2010, \mnras, 407, 1581

\bibitem[{{Silich} {et~al.}(2005){Silich}, {Tenorio-Tagle}, \&
  {A{\~n}orve-Zeferino}}]{silich05}
{Silich}, S., {Tenorio-Tagle}, G., \& {A{\~n}orve-Zeferino}, G.~A. 2005, \apj,
  635, 1116

\bibitem[{{Silich} {et~al.}(2001){Silich}, {Tenorio-Tagle}, {Terlevich},
  {Terlevich}, \& {Netzer}}]{silich01}
{Silich}, S.~A., {Tenorio-Tagle}, G., {Terlevich}, R., {Terlevich}, E., \&
  {Netzer}, H. 2001, \mnras, 324, 191

\bibitem[{{Smith} {et~al.}(2012){Smith}, {Foster}, \& {Brickhouse}}]{smith12}
{Smith}, R.~K., {Foster}, A.~R., \& {Brickhouse}, N.~S. 2012, Astronomische
  Nachrichten, 333, 301

\bibitem[{{Smith} {et~al.}(2017){Smith}, {Abraham}, {Allured}, {Bautz},
  {Bookbinder}, {Bregman}, {Brenneman}, {Brickhouse}, {Burrows}, {Burwitz},
  {Cheimets}, {Costantini}, {Dawson}, {DeRoo}, {Falcone}, {Foster}, {Gallo},
  {Grant}, {G{\"u}nther}, {Heilmann}, {Hertz}, {Hine}, {Huenemoerder},
  {Kaastra}, {Kreykenbohm}, {Madsen}, {McEntaffer}, {Miller}, {Miller},
  {Morse}, {Mushotzky}, {Nandra}, {Nowak}, {Paerels}, {Petre}, {Poppenhaeger},
  {Ptak}, {Reid}, {Sanders}, {Schattenburg}, {Schulz}, {Smale}, {Temi},
  {Valencic}, {Walker}, {Willingale}, {Wilms}, \& {Wolk}}]{smith17}
{Smith}, R.~K., {et~al.} 2017, in Society of Photo-Optical Instrumentation
  Engineers (SPIE) Conference Series, Vol. 10397, Society of Photo-Optical
  Instrumentation Engineers (SPIE) Conference Series, 103970Q

\bibitem[{{Snowden} {et~al.}(2004){Snowden}, {Collier}, \& {Kuntz}}]{snowden04}
{Snowden}, S.~L., {Collier}, M.~R., \& {Kuntz}, K.~D. 2004, \apj, 610, 1182

\bibitem[{{Strickland} {et~al.}(2004){Strickland}, {Heckman}, {Colbert},
  {Hoopes}, \& {Weaver}}]{strickland04}
{Strickland}, D.~K., {Heckman}, T.~M., {Colbert}, E.~J.~M., {Hoopes}, C.~G., \&
  {Weaver}, K.~A. 2004, \apjs, 151, 193

\bibitem[{{Su} \& {Irwin}(2013)}]{su13}
{Su}, Y., \& {Irwin}, J.~A. 2013, \apj, 766, 61

\bibitem[{{Tang} {et~al.}(2009){Tang}, {Wang}, {Mac Low}, \& {Joung}}]{tang09}
{Tang}, S., {Wang}, Q.~D., {Mac Low}, M.-M., \& {Joung}, M.~R. 2009, \mnras,
  398, 1468

\bibitem[{{Temple} {et~al.}(2005){Temple}, {Raychaudhury}, \&
  {Stevens}}]{temple05}
{Temple}, R.~F., {Raychaudhury}, S., \& {Stevens}, I.~R. 2005, \mnras, 362, 581

\bibitem[{{Tenorio-Tagle}(2002)}]{tenorio-tagle02}
{Tenorio-Tagle}, G. 2002, in Revista Mexicana de Astronomia y Astrofisica
  Conference Series, Vol.~12, Revista Mexicana de Astronomia y Astrofisica
  Conference Series, ed. W.~J. {Henney}, J.~{Franco}, \& M.~{Martos}, 50--55

\bibitem[{{T{\"u}llmann} {et~al.}(2006){T{\"u}llmann}, {Pietsch}, {Rossa},
  {Breitschwerdt}, \& {Dettmar}}]{tullmann06}
{T{\"u}llmann}, R., {Pietsch}, W., {Rossa}, J., {Breitschwerdt}, D., \&
  {Dettmar}, R.-J. 2006, \aap, 448, 43

\bibitem[{{Uhlig} {et~al.}(2012){Uhlig}, {Pfrommer}, {Sharma}, {Nath},
  {En{\ss}lin}, \& {Springel}}]{uhlig12}
{Uhlig}, M., {Pfrommer}, C., {Sharma}, M., {Nath}, B.~B., {En{\ss}lin}, T.~A.,
  \& {Springel}, V. 2012, \mnras, 423, 2374

\bibitem[{{Veilleux} {et~al.}(2005){Veilleux}, {Cecil}, \&
  {Bland-Hawthorn}}]{veilleux05}
{Veilleux}, S., {Cecil}, G., \& {Bland-Hawthorn}, J. 2005, \araa, 43, 769

\bibitem[{{Wang}(2010)}]{wang10}
{Wang}, Q.~D. 2010, Proceedings of the National Academy of Science, 107, 7168

\bibitem[{{Weaver} {et~al.}(1977){Weaver}, {McCray}, {Castor}, {Shapiro}, \&
  {Moore}}]{weaver77}
{Weaver}, R., {McCray}, R., {Castor}, J., {Shapiro}, P., \& {Moore}, R. 1977,
  \apj, 218, 377

\bibitem[{{White} \& {Frenk}(1991)}]{white91}
{White}, S.~D.~M., \& {Frenk}, C.~S. 1991, \apj, 379, 52

\bibitem[{{White} \& {Rees}(1978)}]{white78}
{White}, S.~D.~M., \& {Rees}, M.~J. 1978, \mnras, 183, 341

\bibitem[{{Wiegert} {et~al.}(2015){Wiegert}, {Irwin}, {Miskolczi}, {Schmidt},
  {Mora}, {Damas-Segovia}, {Stein}, {English}, {Rand}, {Santistevan},
  {Walterbos}, {Krause}, {Beck}, {Dettmar}, {Kepley}, {Wezgowiec}, {Wang},
  {Heald}, {Li}, {MacGregor}, {Johnson}, {Strong}, {DeSouza}, \&
  {Porter}}]{wiegert15}
{Wiegert}, T., {et~al.} 2015, \aj, 150, 81

\bibitem[{{Wilms} {et~al.}(2000){Wilms}, {Allen}, \& {McCray}}]{wilms00}
{Wilms}, J., {Allen}, A., \& {McCray}, R. 2000, \apj, 542, 914

\bibitem[{{Zhang} \& {Davis}(2017)}]{zhang_dong17}
{Zhang}, D., \& {Davis}, S.~W. 2017, \apj, 839, 54

\bibitem[{{Zhang} \& {Thompson}(2012)}]{zhang_dong12}
{Zhang}, D., \& {Thompson}, T.~A. 2012, \mnras, 424, 1170

\bibitem[{{Zhang} {et~al.}(2014){Zhang}, {Wang}, {Ji}, {Smith}, {Foster}, \&
  {Zhou}}]{zhang14}
{Zhang}, S., {Wang}, Q.~D., {Ji}, L., {Smith}, R.~K., {Foster}, A.~R., \&
  {Zhou}, X. 2014, \apj, 794, 61

\end{thebibliography}

\end{document}